\pdfoutput=1
\documentclass[twocolumn,
superscriptaddress,aps,
prd,
preprintnumbers,amsmath,amssymb,nofootinbib,
]{revtex4-1}

\usepackage{placeins}
\usepackage{amsmath}
\usepackage{amssymb}
\usepackage{amsfonts}
\usepackage{graphicx}
\usepackage[usenames,dvipsnames,svgnames,table]{xcolor}
\usepackage{xfrac}
\usepackage{comment}
\usepackage{pifont}
\usepackage{physics}
\usepackage{hyperref}
\usepackage{bm}
\usepackage{enumitem}

\definecolor{rossoferrari}{HTML}{D9073D}
\definecolor{mediumblue}{HTML}{0000CD}
\definecolor{forestgreen}{HTML}{228B22}
\definecolor{desy_blue}{HTML}{009EE2}
\definecolor{desy_orange}{HTML}{FD8800}
\definecolor{peera_green}{HTML}{008B8B}
\definecolor{peera_orange}{HTML}{B22222}
\definecolor{light_pink}{rgb}{1,0.4,0.4}
\definecolor{light_blue}{rgb}{0.284602,0.317763,0.963947}
\definecolor{peera_col}{RGB}{240, 94, 28}
\definecolor{blue_col}{RGB}{0,92,175}
\definecolor{red_col}{RGB}{203,64,66}
\usepackage{hyperref}
\hypersetup{linktocpage,
    colorlinks=true,
    linkcolor=red_col,
    filecolor=Mahogany,      
    urlcolor=blue_col,
    citecolor=blue_col,
}

\newcommand{\GeV}{\,\mathrm{GeV}}

\def\Mpl{M_\text{Pl}}

\begin{document}


\title{
Constraining Postinflationary Axions with Pulsar Timing Arrays}

\author{G{\'e}raldine Servant} 
\email{geraldine.servant@desy.de}
\affiliation{Deutsches Elektronen-Synchrotron DESY, Notkestr. 85, 22607 Hamburg, Germany}
\affiliation{II. Institute of Theoretical Physics, Universit\"{a}t  Hamburg, 22761, Hamburg, Germany}

\author{Peera Simakachorn}
\email{peera.simakachorn@ific.uv.es}
\affiliation{Instituto de F\'isica Corpuscular (IFIC), Universitat de Val\`{e}ncia-CSIC,\\C/ Catedrático José Beltrán 2, E-46980, Paterna, Spain}

\preprint{DESY-23-094 }

\date{\today}


\begin{abstract}
\noindent

Models that produce Axion-Like-Particles (ALP) after cosmological inflation due to spontaneous $U(1)$ symmetry breaking also produce cosmic string networks.
Those axionic strings lose energy
through gravitational wave emission during the whole cosmological history,  generating a stochastic background of gravitational waves that spans many decades in frequency.  We can therefore constrain the axion decay constant and axion mass from limits on the gravitational wave spectrum and compatibility with dark matter abundance as well as dark radiation. 
We derive such limits from analyzing the most recent NANOGrav data from Pulsar Timing Arrays (PTA).
The limits are similar to the  $N_{\rm eff}$ bounds on dark radiation
for ALP masses $m_a \lesssim 10^{-22}$ eV. On the other hand, for heavy ALPs with  $m_a\gtrsim 0.1$ GeV and $N_{\rm DW}\neq 1$,  new regions of parameter space can be probed by PTA data due to the dominant  Domain-Wall contribution to the gravitational wave background.

\end{abstract}


\maketitle

\section{Introduction}
Pulsar Timing Arrays (PTA) offer a new window to observe the Universe through gravitational waves (GW) in the nano-Hertz frequency range \cite{NANOGrav:2023gor, Antoniadis:2023ott, Reardon:2023gzh, Xu:2023wog, NANOGrav:2023hvm,Antoniadis:2023xlr}.
A potential source of GWs at these frequencies is a population of supermassive black-hole binaries (SMBHBs) in the local universe \cite{NANOGrav:2023hfp, Antoniadis:2023xlr}.
Besides, cosmic strings, which may have been produced in the early Universe during a spontaneous  $U(1)$ symmetry-breaking event \cite{Kibble:1976sj, Kibble:1980mv, Hindmarsh:1994re, Vilenkin:2000jqa}, generate a stochastic gravitational-wave background (SGWB) down to these low frequencies as part of a vast spectrum spanning many decades in frequency; see \cite{Gouttenoire:2019kij, Auclair:2019wcv} for recent reviews.
In fact, given the very wide and nearly scale-invariant GW spectrum from cosmic strings, the PTA limits are very relevant to anticipate the prospects for probing a cosmic-string GW signal at LISA \cite{LISACosmologyWorkingGroup:2022jok} or Einstein Telescope \cite{Maggiore:2019uih}. 
Cosmic strings can either be  \emph{local} or \emph{global} depending on whether the spontaneously broken symmetry is a gauge or global ~$U(1)$.
Models of local strings have been confronted to PTA data in \cite{Ellis:2020ena, Blasi:2020mfx, Buchmuller:2020lbh, Samanta:2020cdk}, and most recently to the 15-year NANOGrav (NG15) data in \cite{NANOGrav:2023hvm} and the EPTA data release 2 \cite{EPTA:2023hof, Antoniadis:2023xlr}.

This paper focuses instead on GW from global strings \cite{Chang:2019mza, Chang:2021afa, Gouttenoire:2019kij, Gorghetto:2021fsn, Ramberg:2019dgi, Ramberg:2020oct}, which were not analyzed in \cite{NANOGrav:2023hvm}.
Many Standard-Model extensions feature such additional global $U(1)$ symmetry that gets spontaneously broken by the vacuum expectation value of a complex scalar field, thus delivering a Nambu-Goldstone Boson. A famous example is the Peccei-Quinn $U(1)$ symmetry advocated to solve the strong CP problem and its associated axion particle \cite{Peccei:1977hh, Peccei:1977ur, Weinberg:1977ma, Wilczek:1977pj}.
Because the $U(1)$ symmetry gets also broken explicitly at later times, the axion acquires a mass. At that moment, domain walls can also populate the Universe \cite{Sikivie:1982qv}.

This paper considers this broad class of models of so-called \emph{Axion-Like-Particles} (ALPs) with mass $m_a$ and decay constant $f_a$, corresponding to the energy scale of spontaneous symmetry breaking.
 If the cosmic-string and domain-wall formations happen before inflation, those are diluted away.
 On the other hand, if the $U(1)$ is broken at the end or after inflation (in this case, the ALP is dubbed {\it postinflationary}), cosmic strings give rise to a population of loops that generate a SGWB throughout the cosmic history. At the same time, they also generate axion particles \cite{Davis:1986xc, Davis:1989nj, Dabholkar:1989ju, Battye:1993jv, Gorghetto:2018myk, Gorghetto:2020qws, Buschmann:2021sdq}, while domain walls bring an additional contribution to the GW spectrum \cite{Lyth:1991bb, Nagasawa:1994qu, Chang:1998tb,  Hiramatsu:2012gg, Hiramatsu:2012sc, Gelmini:2021yzu, Gelmini:2022nim, Gelmini:2023ngs}.

We aim to use the most recent limits on the SGWB from 
NG15 data set to derive independent bounds on the parameter space of postinflationary ALPs.
Given that a GW signal has been observed \cite{NANOGrav:2023gor}, any further improved sensitivity from future PTA observatories will not enable pushing down the constraints. Therefore, the PTA constraints presented in this paper on the axion mass and decay constant are not expected to change by more than a factor of a few from future PTA experiments. On the other hand, future GW experiments operating in other frequency ranges will serve as complementary probes to PTA.

Our approach is the following. 
We analyze the recent NG15 data via the code \texttt{PTArcade} \cite{andreamitridate2023, Mitridate:2023oar}, first considering 
the two SGWB from global cosmic strings and domain walls without the astrophysical background. 
We compare the interpretation of data in terms of SMBHBs and of global cosmic strings and domain walls by calculating the Bayes Factor (BF).
Next, we set constraints on the new physics contribution, leading to a SGWB that is too strong and conflicts with the data.
The results of the best fit and the constraints on the SGWB from domain walls have been presented in the recent analysis with NG15 data by the NANOGrav collaboration  \cite{NANOGrav:2023hvm}.
Regarding the analysis of previous data release, Refs.~\cite{Ferreira:2022zzo, Bian:2022qbh, Madge:2023cak} fitted the domain-wall and/or global-string signal to the PPTA second data release (DR2) or IPTA DR2 and/or NANOGrav 12.5-year data, however did not derive the exclusion region. 
(See Sec.~\ref{sec:data_analysis} for more details on the best fit and constraint.)
We further translate these bounds into constraints in the ALP parameter space. 
In addition, this work presents a similar analysis (determining best fits and setting constraints) for global strings for the first time with NG15.

Sec.~\ref{sec:model} of this \emph{paper} summarizes the postinflationary axion scenarios and their corresponding GW signals, separated into two cases: either cosmic-string or domain-wall SGWB dominates.
In Sec.~\ref{sec:data_analysis}, we confront these cases with the NG15 data and derive, for each case, the constraints on axion parameter space, illustrated in Fig.~\ref{fig:ma_fa}. We conclude in Sec.~\ref{sec:conclude}. Supplemental material contains miscellaneous details, such as the priors for analysis, the case assuming no astrophysical background, and the result for the global strings in the $m_a \to 0$ limit.

\section{Postinflationary axion and its gravitational waves}
\label{sec:model}
The ALP can be defined as the angular mode $\theta$ of a  complex scalar field $\Phi \equiv \phi \exp(i\theta)$ with $\phi$ the radial partner. It has the Lagrangian density, $\mathcal{L} = \frac{1}{2}\partial_\mu \Phi^* \partial^\mu \Phi -  V(\Phi) - V_{\rm c}$ with $V_{\rm c}$ the  correction responsible for $U(1)$ symmetry restoration and trapping $\Phi \to 0$ at early times. The potential has three terms: 
    \begin{align}
        V(\Phi) = \underbrace{\frac{\lambda}{2}(\phi^2 - f_a^2)^2}_{\rm cosmic ~strings} + \underbrace{\frac{m_a^2 f_a^2}{N_{\rm DW}^2}\left[1 - \cos\left(N_{\rm DW} \theta \right)\right]}_{\rm domain~walls} + V_{\rm bias}, \nonumber
    \end{align}
    where $f_a$ is the vacuum expectation value of the field, $m_a \equiv m_a(T)$ is the axion mass as a function of the Universe's temperature $T$, $N_{\rm DW}$ is the number of domain walls, and $V_{\rm bias}$ is some further explicit $U(1)$ breaking term. The first term is responsible for $U(1)$ spontaneous breaking, while the second and third terms explicitly break the $U(1)$.
These three terms are ranked according to their associated energy scales (large to small) corresponding to their sequences in defect formations: from cosmic strings to domain walls and then their decays.

During inflation, the complex scalar field is driven to the minimum of the potential $V(\Phi)$ if $V_c \ll V(\Phi)$.
Quantum fluctuations along the axion direction due to the de Sitter temperature 
${\cal O}(H_{\inf})$
can generate a positive quadratic term in the potential and restore the $U(1)$ symmetry, which gets eventually broken at the end of inflation, leading to cosmic strings if $H_{\inf}/(2 \pi f_a) \gtrsim 1$ \cite{Bunch:1978yq, Linde:1983mro, Starobinsky:1994bd}. However, the current CMB bound \cite{Planck:2018jri} on the inflationary scale $H_{\inf} < 6.1 \times 10^{13} \, {\rm GeV}$ implies that $f_a$ is too small to generate an observable cosmic-string SGWB. Still, there are several other ways in which 
$U(1)$ can get broken after inflation even for large $f_a$:
\emph{i)} 
A large and positive effective $\phi$-mass can be generated by coupling $\phi$ to the inflaton $\chi$ (\emph{e.g.,} $\mathcal{L} \supset \chi^2 \phi^2$) which, for large $\chi$, traps $\phi \to 0$ during 
inflation\footnote{As the inflaton field value relates to the Hubble parameter, this mass is called \emph{Hubble-induced} mass.}.
\emph{ii)} $\phi$ could couple to a thermal (SM or secluded) plasma of temperature $T$ that would generate a large thermal $V_c$ correction, restoring the $U(1)$\footnote{For example, the KSVZ-type of interaction couples $\phi$ to a fermion $\psi$  charged under some gauge symmetry with $A_\mu$: $\mathcal{L} \supset y \phi \bar{\psi} \psi + {\rm h.c.} + g \bar{\psi} \gamma^\mu \psi A_\mu$, that can generate thermal corrections: $V_c = y^2 T^2 \phi^2$ for $y \phi < T$ and $V_c = g^4 T^4 \ln(y^2\phi^2/T^2)$ for $y \phi \gtrsim T$ \cite{Mukaida:2012qn, Mukaida:2012bz}. When $V_c > \lambda f_a^4$, the $\phi$-field is trapped at the origin at temperature $T \gtrsim \sqrt{\lambda}f_a/y$ for $y f_a < T$ and $T \gtrsim \lambda^{1/4} f_a/g$ for $y f_a > T$.
For couplings of order unity, $f_a < T < T_{\rm max} \simeq 6.57 \times 10^{15} {\GeV}$ is the maximum reheating temperature bounded by the inflationary scale and assuming instantaneous reheating. Nonetheless, if $\lambda$ is small (corresponding to a small radial-mode mass), the bound can be weakened.}. \emph{iii)} Lastly, non-perturbative processes, such as preheating, could also lead to $U(1)$ 
restoration after inflation \cite{Kofman:1995fi, Tkachev:1995md, Kasuya:1997ha, Kasuya:1998td, Tkachev:1998dc}.

When $V_{\rm c}$ drops, the first term of $V(\Phi)$ breaks spontaneously the $U(1)$ symmetry at energy scale $f_a$, leading to the network formation of line-like defects or \emph{cosmic strings} with tension $\mu = \pi f_a^2 \log(\lambda^{1/2}f_a/H)$ \cite{Vilenkin:2000jqa}. As $U(1)$ symmetry is approximately conserved when the axion mass is negligible, the cosmic strings survive for long and evolve into the \emph{scaling regime} by chopping-off loops \cite{Kibble:1984hp, Albrecht:1984xv, Bennett:1987vf, Bennett:1989ak, Albrecht:1989mk, Allen:1990tv, Martins:2000cs, Ringeval:2005kr, Vanchurin:2005pa, Martins:2005es, Olum:2006ix, Blanco-Pillado:2011egf, Figueroa:2012kw, Martins:2016ois}. Loops are continuously produced and emit GW and axion particles throughout cosmic history.
The resulting  GW signal corresponds to a SGWB entirely characterized by its frequency power spectrum. The latter is commonly expressed as the GW fraction of the total  energy density of the Universe $h^2 \Omega_{\rm GW}(f_{\rm GW})$.

A loop population  produced at temperature $T$  quickly decays into GW of frequency \cite{Gouttenoire:2019kij},
\begin{align}
f_{\rm GW}^{\rm CS} (T) \simeq 63 ~ {\rm nHz} \left(\frac{\alpha}{0.1}\right) \left(\frac{T}{10 \rm MeV}\right) \left[\frac{g_*(T)}{10.75}\right]^{\frac{1}{4}},
\label{eq:freq_cs}
\end{align}
where $\alpha \sim \mathcal{O}(0.1)$ is the typical loop size in units of the Hubble horizon $1/H$.
If the network of cosmic strings is stable until late times, \emph{i.e.,} in the limit $m_a \to 0$, its SGWB is characterized by \cite{Gouttenoire:2019kij, Gouttenoire:2021jhk},
\begin{align}
h^2 \Omega_{\rm GW}^{\rm CS} (f_{\rm GW})  \simeq 1.3 \times 10^{-9} \left(\frac{f_a}{3 \times 10^{15} \, \rm GeV}\right)^4 \times \nonumber\\
\times \mathcal{G}(T(f_{\rm GW}))\left[\frac{\mathcal{D}(f_a, f_{\rm GW})}{94.9}\right]^3 \left[\frac{C_{\rm eff}(f_{\rm GW})}{2.24}\right],
\label{eq:amp_cs}
\end{align}
where $\mathcal{G}(T) \equiv [g_*(T)/g_*(T_0)][g_{*s}(T_0)/g_{*s}(T)]^{4/3}$ with $T_0$ the temperature of the Universe today. The logarithmic correction is defined by
\begin{align}
    \mathcal{D}(f_a, f_{\rm GW})=\log\left[1.7 \cdot 10^{41} \left(\frac{f_a}{3 \cdot 10^{15} \, \rm GeV}\right) \left(\frac{10 \, \rm nHz}{f_{\rm GW}}\right)^2\right],
\end{align}
and $C_{\rm eff}(f_{\rm GW})$ is the loop-production efficiency which also receives a small log correction originated from axion production \cite{Gouttenoire:2019kij}.
$g_{*}$ and $g_{*s}$ measure the number of relativistic degrees of freedom in the energy and entropy densities, respectively.
Note that the exponent `3' of the log-dependent term $\mathcal{D}$ is still under debate \cite{Gorghetto:2018myk,Gorghetto:2020qws, Kawasaki:2018bzv, Vaquero:2018tib, Klaer:2017qhr, Klaer:2019fxc, Hindmarsh:2019csc, Figueroa:2020lvo, Buschmann:2019icd, Buschmann:2021sdq, Gorghetto:2021fsn, Hindmarsh:2021vih}.  E.g., some numerical simulations find the scaling network leading to the exponent `3' \cite{Hindmarsh:2021vih}, while  the non-scaling one leads to the exponent `4' \cite{Gorghetto:2018myk,Gorghetto:2020qws,Gorghetto:2021fsn}.
From Eq.~\eqref{eq:amp_cs}, the uncertainty in $\Omega_{\rm GW}$ due to a factor of $\mathcal{D} \sim \mathcal{O}(100)$ leads to the uncertainty in the constraint on $f_a$ by $\sim 10^{1/2}$.
Moreover, the recent debate on the GW-emission power from a single loop in different numerical simulations is open \cite{Gorghetto:2021fsn, Baeza-Ballesteros:2023say}.
 This work uses the semi-analytic result, \emph{e.g.,} in   \cite{Chang:2019mza, Chang:2021afa, Gouttenoire:2019kij}, which predicts $\Omega_{\rm GW}$ that is weaker than \cite{Gorghetto:2021fsn} and stronger than \cite{Baeza-Ballesteros:2023say}.
   
As the Universe cools, the axion mass develops due to non-perturbative effects (like  strong confinement in the case of the QCD axion). 
The second term in $V(\Phi)$  breaks explicitly the $U(1)$ discretely, leading to sheet-liked defects or \emph{domain walls}, attached to the cosmic strings. 
The domain wall is characterized by its surface tension $\sigma \simeq 8 m_a f_a^2/N_{\rm DW}^2$ \cite{Hiramatsu:2012sc}. The axion field starts to feel the presence of the walls when $3 H \simeq m_a$.
The domain-wall network can be stable or unstable depending on the number of domain walls attached to a string.
The value of $N_{\rm DW}$ is very UV-model-dependent.  It can be linked to the discrete symmetry $Z_{N_{\rm DW}}$ \cite{Dine:1981rt, Zhitnitsky:1980tq, Kim:1986ax} that remains after the confinement of the gauge group that breaks the global $U(1)$  symmetry explicitly and generates the axion mass.
This occurs at the scale $\Lambda \simeq \sqrt{m_a F_a}$, where $F_a = f_a / N_{\rm DW} $, that is when 
the domain walls are generated, attaching to the existing cosmic strings.

For $N_{\rm DW}>1$, the string-wall system is stable and long-lived. Its decay may be induced by $V_{\rm bias}$, the biased term \cite{Vilenkin:1981zs, Gelmini:1988sf, Larsson:1996sp}, which could be of QCD origin \cite{Sikivie:1982qv, Chang:1998tb}. This decay is desirable to prevent DW from dominating the energy density of the Universe at late times. $V_{\rm bias}$ is therefore an additional free parameter beyond $m_a$ and $f_a$ that enters the GW prediction in the case where $N_{\rm DW}> 1$.

\begin{figure}[t!]
	\centering
	\includegraphics[width=\linewidth]{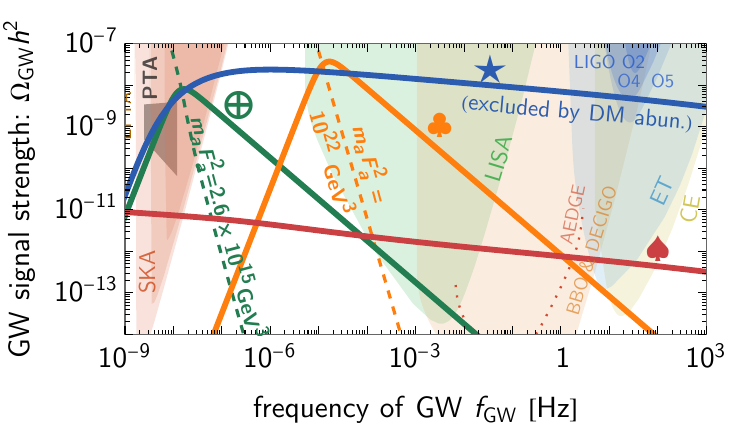}
 \vspace{-2em}
\caption{ Solid curves show the SGWB spectra from axionic strings for $N_{\rm DW}=1$ \{\textcolor{RoyalBlue}{blue-{\Large $\star$}}, \textcolor{BrickRed}{red-$\spadesuit$}\} and domain walls for $N_{\rm DW} > 1$ \{\textcolor{ForestGreen}{green-$\bigoplus$}, \textcolor{BurntOrange}{orange-$\clubsuit$}\}. The symbols correspond to the benchmark points in the axion parameter space in Fig.~\ref{fig:ma_fa} (with $T_\star= \{128 \, {\rm MeV}, 10^2 \, {\rm GeV}$\} for \{$\bigoplus$, $\clubsuit$\}). The best-fitted spectra to the PTA data are the blue-{\Large $\star$} curve for global strings, corresponding to $\{f_a,m_a\} \simeq \{9.9 \times 10^{15} ~ {\rm GeV}, 4.8 \times 10^{-15} ~ {\rm eV}\}$ which is excluded by the axion dark matter abundance [see Eq.~\eqref{eq:dm_axion_string}], and the green-$\bigoplus$ curve for domain walls (with $m_a F_a^2 = 2.6 \times 10^{15} ~ {\rm GeV}^3$). The power-law integrated sensitivity curves of GW experiments \cite{Janssen:2014dka, Lentati:2015qwp, Desvignes:2016yex, NANOGRAV:2018hou, Weltman:2018zrl, LISA:2017pwj, LISACosmologyWorkingGroup:2022jok, Yagi:2011wg, AEDGE:2019nxb, KAGRA:2013rdx, LIGOScientific:2014qfs, LIGOScientific:2019vic, Hild:2010id, Punturo:2010zz, LIGOScientific:2016wof} are taken from \cite{Gouttenoire:2019kij, Breitbach:2018ddu}. For fixed $\{m_a, f_a\}$ values, the peak of the DW-GW spectrum moves along the dashed line as $T_\star$ varies; see Eq.~\eqref{eq:peak_position}.}
    \label{fig:spectrum_gw}
\end{figure}

\subsection{Case i) $N_{\rm DW} = 1$}
If only one domain wall is attached to a string, \emph{i.e.,} $N_{\rm DW} = 1$, the string-wall system quickly annihilates due to DW tension when\footnote{The string tension loses against the DW surface tension at time $t_{\rm dec}$ defined by \cite{Vilenkin:1982ks}
   $ F_{\rm str} \sim \mu/R_{\rm dec} \simeq \sigma ~ \Rightarrow ~ R_{\rm dec} \sim H^{-1}(t_{\rm dec}) \sim \mu/\sigma \sim m_a^{-1} $
where $R$ is the string curvature, assumed to be of Hubble size.} $H(T_{\rm dec}) \simeq m_a$ \cite{Hiramatsu:2012gg}.
The cosmic string SGWB features an IR cut-off corresponding to the temperature 
\begin{align}
T_{\rm dec} \simeq 1.6 ~{\rm MeV} \left[\frac{10.75}{g_*(T_{\rm dec})}\right]^{\frac{1}{4}} \left(\frac{m_a}{10^{-15} ~ {\rm eV}}\right)^{\frac{1}{2}},
\label{eq:Tdec_ma}
\end{align}
associated with the frequency,
\begin{align}
f_{\rm GW}^{\rm CS} (m_a) \simeq 9.4 ~ {\rm nHz} \left(\frac{\alpha}{0.1}\right)\left(\frac{m_a}{10^{-15} \rm eV}\right)^{\frac{1}{2}}.
\label{eq:freq_cs_axion_mass}
\end{align}
The cut-off position -- frequency and amplitude -- can be estimated with Eqs.~\eqref{eq:amp_cs}--\eqref{eq:freq_cs_axion_mass}.
At $f_{\rm GW} < f_{\rm GW}^{\rm CS} (T_{\rm dec})$, the spectrum scales as $\Omega_{\rm GW} \propto f_{\rm GW}^{3}$ due to causality.
Note that for $m_a \ll 10^{-16}$ eV, the cut-off sits below nHz frequencies, and within the PTA window, we recover the same GW spectrum as the one in the limit $m_a \to 0$.
Our analysis applies the numerical templates of the global-string SGWB -- covering the ranges of $f_a$ and $T_{\rm dec}$ priors. We calculated these templates numerically by solving the string-network evolution via the \emph{velocity-dependent one-scale} model \cite{Martins:1996jp, Martins:2000cs, Sousa:2013aaa, Sousa:2014gka, Correia:2019bdl}, shutting off the loop production after $H = m_a$, and calculating the SGWB following Ref.~\cite{Gouttenoire:2019kij}.

\begin{figure*}[t]
	\centering
	\includegraphics[height=0.28\linewidth]{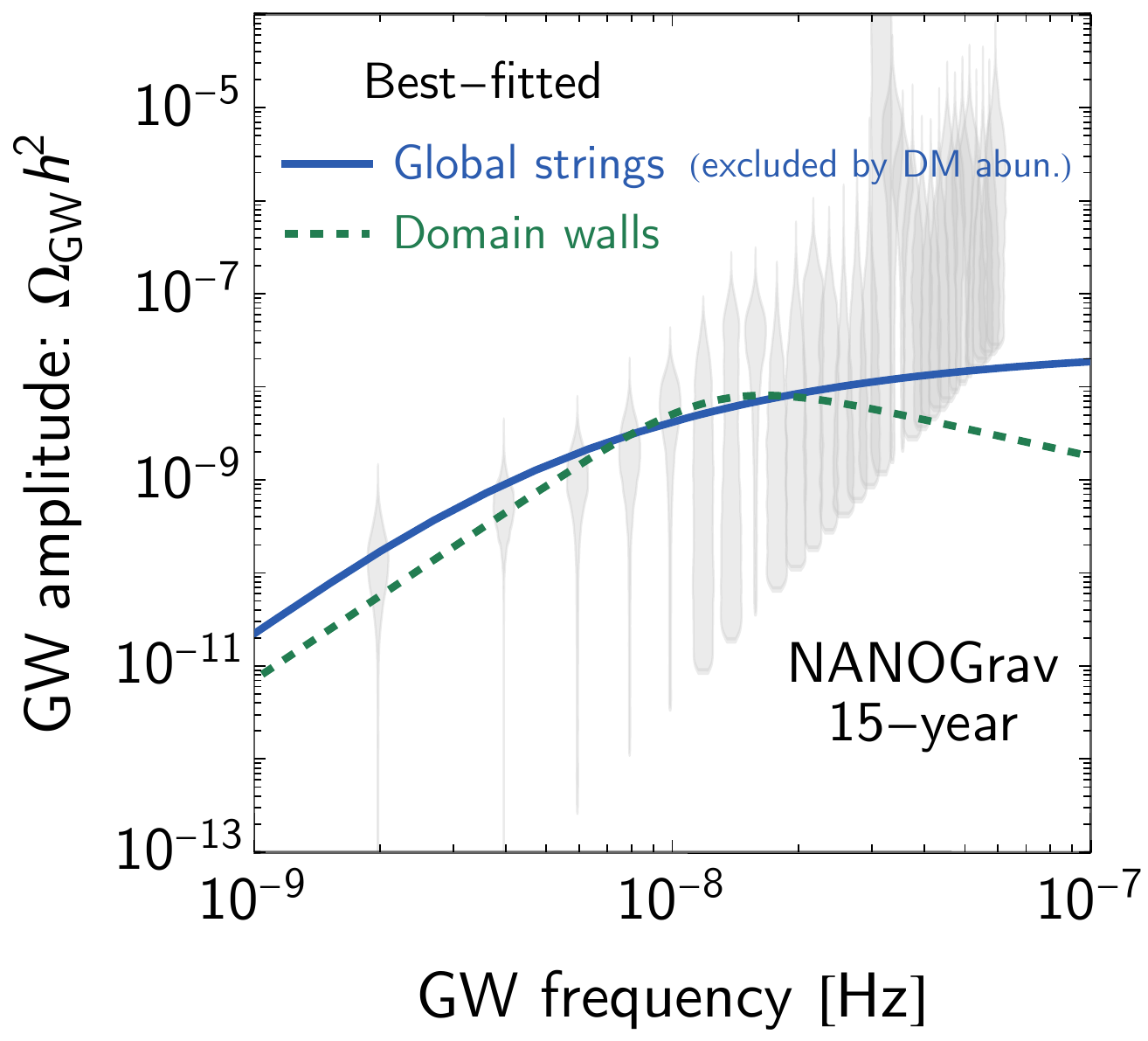} \hspace{-1mm}\hfill
	\includegraphics[height=0.2775\linewidth]{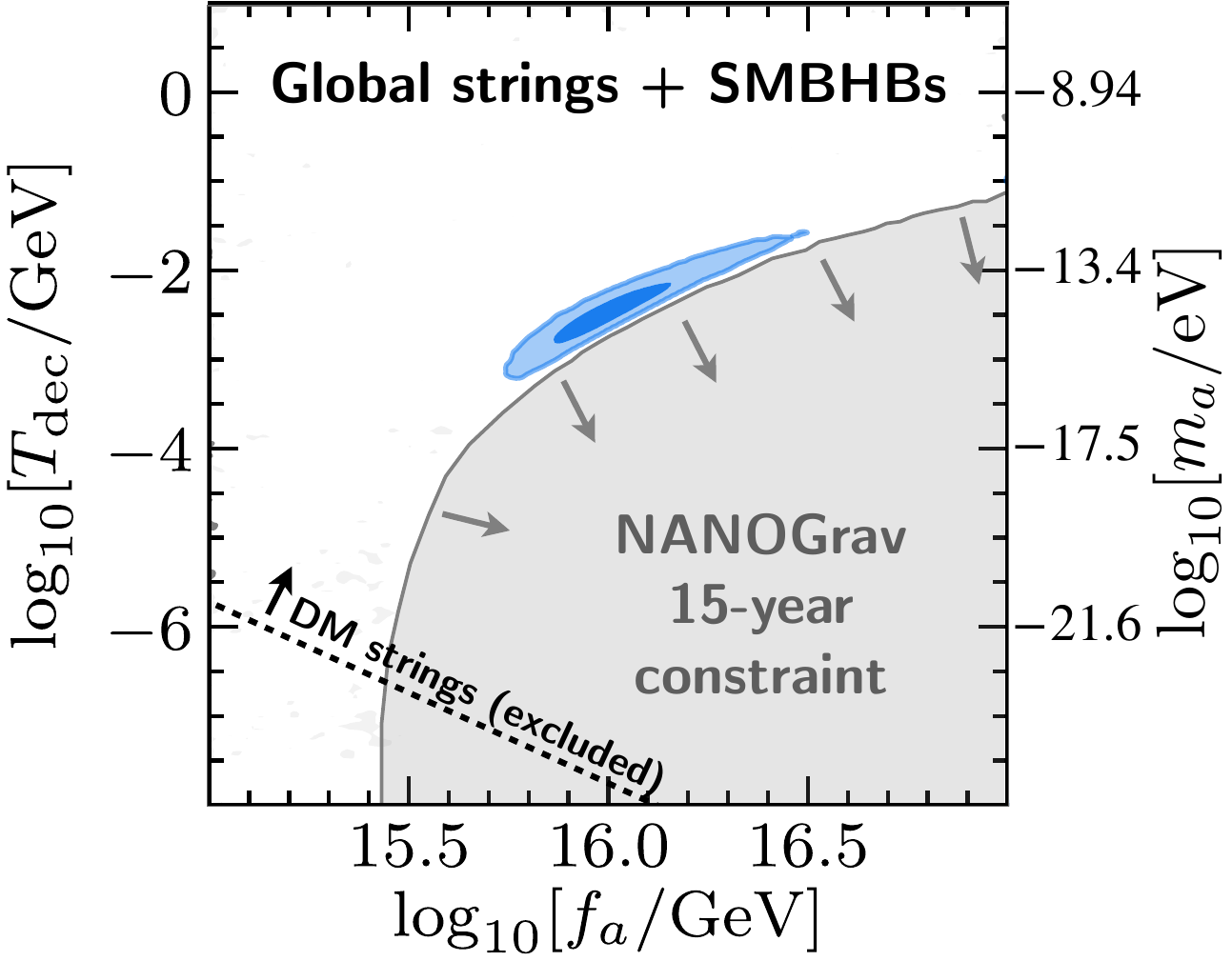} \hspace{-1.25mm}\hfill
	\includegraphics[height=0.2775\linewidth]{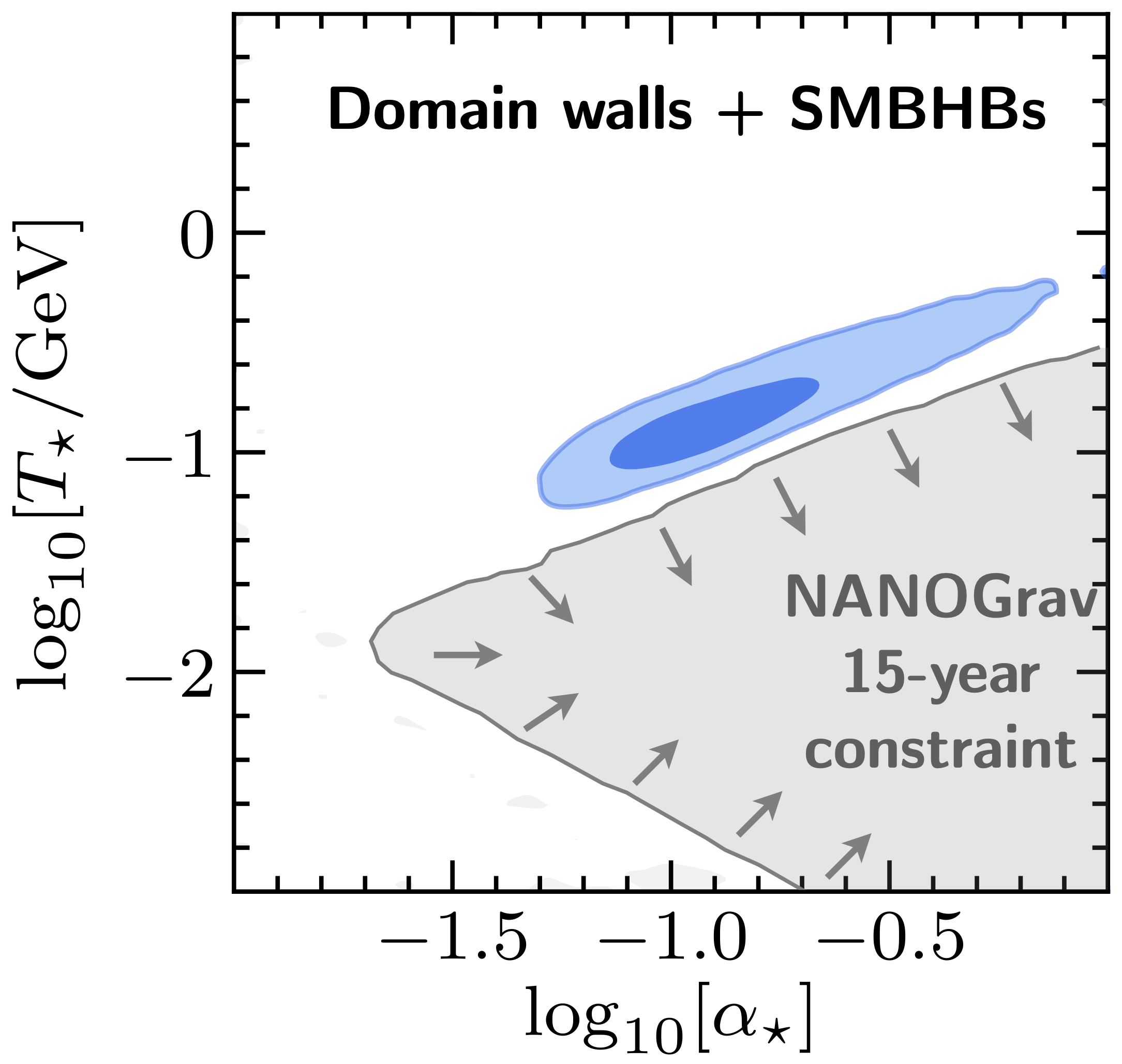} 
 \vspace{-1em}
\caption{\emph{left:}  The SGWB spectra from global strings and domain walls + SMBHBs, providing the best-fits to the PTA data and corresponding to $\{f_a,m_a\} \simeq \{9.9 \times 10^{15} ~ {\rm GeV}, 4.8 \times 10^{-15} ~ {\rm eV}\}$ for global strings and $m_a F_a^2 = 2.6 \times 10^{15} ~ {\rm GeV}^3$ for domain walls (in violins, taken from \cite{NANOGrav:2023hvm}).  \emph{middle and right:} 1$\sigma$ (dark blue) and $2\sigma$ (light blue) regions of the likelihood of the global-string/domain-wall parameters, assuming the template of global-string/domain-wall +  SMBHB backgrounds.
    The gray region is excluded due to too strong GW signals from global strings/domain walls that conflict with PTA data.
    The region above the black dashed line in the middle panel (including the best fit) conflicts with the dark matter abundance [see Eq.~\eqref{eq:dm_axion_string}].
    }
    \label{fig:main_result}
\end{figure*}

\subsection{Case ii) $N_{\rm DW} > 1$} Attached to a string, $N_{\rm DW}$ walls balance among themselves and prevent the system from collapsing at $H \simeq m_a$ \cite{Hiramatsu:2010yn, Hiramatsu:2012gg}.
The domain-wall network later evolves to the  \emph{scaling regime} where there is a constant number of DW per comoving volume $\mathcal{V} \simeq H^{-3}$. The energy density of DW is $\rho_{\rm DW} \simeq \sigma H^{-2}/\mathcal{V} \simeq \sigma H $ and it acts as a long-lasting source of SGWB \cite{Vilenkin:1981zs, Preskill:1991kd,Gleiser:1998na,Hiramatsu:2010yz, Kawasaki:2011vv,Hiramatsu:2013qaa,ZambujalFerreira:2021cte}; \emph{cf.} \cite{Saikawa:2020duz} for a compact review. The network red-shifts slower than the Standard Model (SM) radiation energy density and could dominate the Universe.
The biased term $V_{\rm bias}$ -- describing the potential difference between two consecutive vacua -- explicitly breaks the $U(1)$ symmetry and induces the pressure on one side of the wall \cite{Kibble:1976sj, Vilenkin:1981zs}. Once this pressure overcomes the tension of the wall\footnote{The pressure from $V_{\rm bias}$ is $p_V \sim V_{\rm bias}$, while the wall's tension reads $p_T \sim \sigma H$ assuming the wall of horizon size. The collapse happens when $p_V > p_T$.}, the string-wall system collapses at temperature,
\begin{align}
T_\star \simeq 53 {\rm MeV} \left[\frac{10.75}{g_*(T_*)}\right]^{\frac{1}{4}} \left[\frac{V_{\rm bias}^{\frac{1}{4}}}{10{\rm MeV}}\right]^2 \left[\frac{\rm GeV}{m_a}\right]^{\frac{1}{2}} \left[\frac{10^6 \rm GeV}{f_a/N_{\rm DM}}\right].
\label{eq:Tstar}
\end{align}
The fraction of energy density in DW is maximized at this time and reads,
\begin{align}
\alpha_\star &\equiv \rho_{\rm DW}/\rho_{\rm tot}(T_\star) \simeq \sigma H / (3 \Mpl^2 H^2(T_\star)),\nonumber\\
&\hspace{-3mm} \simeq 4 \times 10^{-4} \left[\frac{10.75}{g_*(T_\star)}\right]^{\frac{1}{2}} \left[\frac{m_a}{\rm GeV}\right] \left[\frac{f_a/N_{\rm DW}}{10^6 \rm GeV}\right]^2  \left[\frac{50 \rm MeV}{T_\star}\right]^2.
\label{eq:DW_alpha}
\end{align}
The energy density emitted in GW is \cite{Hiramatsu:2012sc}
\begin{align}
\rho_{\rm GW}/\rho_{\rm tot} \sim \frac{3}{32\pi}\epsilon\alpha_\star^2 
\end{align}
  where we fix $\epsilon \simeq 0.7$ from numerical simulations \cite{Hiramatsu:2013qaa}. It reaches its maximum at $T_\star$. The spectrum exhibits the broken-power law shape and reads,
\begin{align}
h^2 \Omega_{\rm GW}^{\rm DW} (f_{\rm GW}) \simeq &7.35 \times 10^{-11} \left[\frac{\epsilon}{0.7}\right] \left[\frac{g_*(T_\star)}{10.75}
\right] \left[\frac{10.75}{g_{*s}(T_\star)}
\right]^{\frac{4}{3}} \times \nonumber\\
&\times \left(\frac{\alpha_\star}{0.01}\right)^2 \mathcal{S}\left(\frac{f_{\rm GW}}{f_{\rm p}^{\rm DW}} \right)
\label{eq:GWDWspec}
\end{align}
where the normalized spectral shape is,
\begin{align}
\mathcal{S}(x) = (3 + \beta)^\delta / (\beta x^{-\frac{3}{\delta}} + 3 x^{\frac{\beta}{\delta}})^\delta.
\label{eq:spectralshape}
\end{align}
The $f_{\rm GW}^3$-IR slope is dictated by causality, the UV slope $f_{\rm GW}^\beta$ is model-dependent, and the width of the peak is $\delta$.
The peak frequency corresponds to the DW size, \emph{i.e.,} the horizon size $f_{\rm GW}^\star \sim H_{\star}$ \cite{Hiramatsu:2013qaa}.  Its value today reads, 
\begin{align}
f_{\rm p}^{\rm DW} \simeq 1.14 {\rm nHz} \left[\frac{g_*(T_\star)}{10.75}
\right]^{\frac{1}{2}} 
\left[\frac{10.75}{g_{*s}(T_\star)}
\right]^{\frac{1}{3}} \left[\frac{T_\star}{10  \rm MeV}\right].
\label{eq:freq_dw}
\end{align}
From Eqs.~\eqref{eq:DW_alpha}, \eqref{eq:GWDWspec}, and \eqref{eq:freq_dw}, each value of $m_a f_a^2$ corresponds to a degenerate peak position of the GW spectrum,
\begin{align}
h^2 \Omega_{\rm GW}^{\rm DW} (f_{\rm p}^{\rm DW}) \simeq &1.2 \times 10^{-10} \left[\frac{\epsilon}{0.7}\right] \left[\frac{g_*(T_\star)}{10.75}
\right]^3 \left[\frac{10.75}{g_{*s}(T_\star)}
\right]^{\frac{8}{3}} \times \nonumber\\
&\times \left[\frac{m_a}{\rm GeV}\right]^2 \left[\frac{f_a}{10^6 \rm GeV}\right]^4\left[\frac{\rm nHz}{f_{\rm p}^{\rm DW}}\right]^4,
\label{eq:peak_position}
\end{align}
which are shown as the dashed line in Fig.~\ref{fig:spectrum_gw}.

The DW can decay into axions, which either behave as dark radiation or decay into SM particles. When DW decay into dark radiation, the $\Delta N_{\rm eff}$ puts a bound $\alpha_\star \lesssim 0.06$ \cite{Ferreira:2022zzo}, \emph{i.e.,} the peak of GW spectrum has $h^2\Omega_{\rm GW} \lesssim 10^{-9}$ (which cannot fit the whole 14 bins of NG15 data).
As $\alpha_\star$ controls the amplitude of the GW spectrum \eqref{eq:GWDWspec}, we consider a larger range of $\alpha_\star$, up to $\alpha_\star = 1$ when the energy density of DW starts to dominate the Universe. To get around the $\Delta N_{\rm eff}$ bound, we will therefore consider the case where the axions produced by domain walls eventually decay into SM particles.

\begin{figure*}[t]
	\centering
     \includegraphics[width=\linewidth]{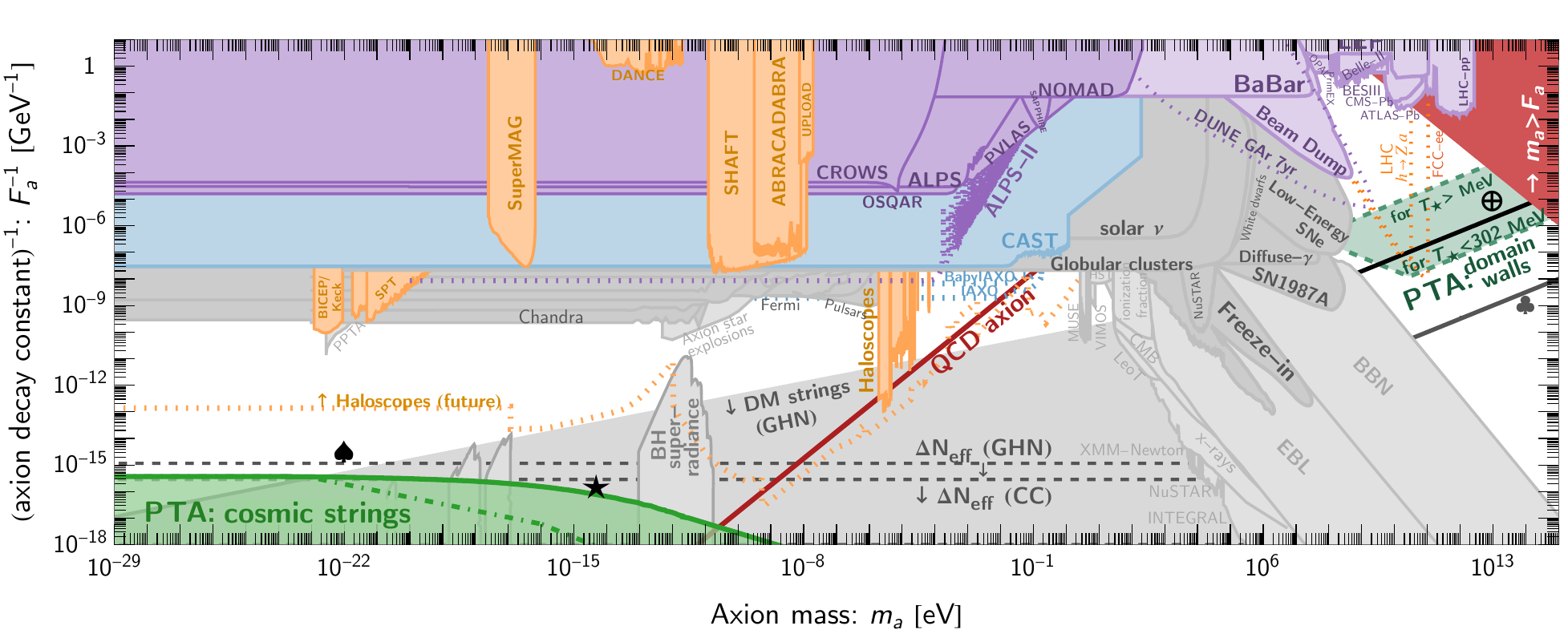}
    \vspace{-2em}
	\caption{PTA limits (in green) on postinflationary axions, compared to existing experimental constraints as compiled from \texttt{AxionLimits} \cite{AxionLimits} and to theoretical bounds: dark radiation overabundance $\Delta N_{\rm eff}$ bound \eqref{eq:neff_axion_string} as dashed horizontal line and ALP overabundance \eqref{eq:dm_axion_string} in the shaded grey region.  $F_a = f_a / N_{\rm DW} $.
The orange dotted lines in the $m_a \gtrsim$ 1 GeV  region are the projections of  future collider experiments, LHC ($h \to Z a$) and FCC ($e^+ e^- \to h a$), obtained from \cite{Bauer:2017ris, Bauer:2018uxu} with the maximally allowed ALP-SM coupling.
 The red region denoted $m_a > F_a$ is where the axion effective field theory is not valid.
 The comparison with experimental bounds uses 
 $g_{\theta\gamma\gamma}=1.02\alpha_{\rm EM}/(2\pi F_a) \approx 2.23 \times  10^{-3}/F_a$ 
 for the relation between the photon coupling and $F_a$, as motivated by KSVZ models \cite{Kim:1979if, Shifman:1979if}. The recent PTA data \cite{NANOGrav:2023gor} excludes the \textcolor{ForestGreen}{\bf green} small-$m_a$ region due to cosmic-string SGWB  ($N_{\rm DW} =1$). It also potentially excludes the high-$m_a$ region due to domain-wall SGWB for $N_{\rm DW} > 1$, depending on the value of $T_\star$. 
 The \textcolor{PineGreen}{\bf other green} band at large $m_a$ is the region that PTA can constrain if $T_\star$ varies in the range MeV $< T_\star < 302$ MeV, as illustrated in Fig.~\ref{fig:largemass_constraint}. The two benchmark points \{{\Large $\star$}, $\spadesuit$\} correspond to cosmic-string SGWB, and the two black benchmark lines \{$\bigoplus$, $\clubsuit$\} correspond to the domain-wall SGWB, whose spectra are shown in Fig.~\ref{fig:spectrum_gw}. The green dot-dashed line is explained in App.\ref{app:EMD_GW_cosmic_string}.
 }
	\label{fig:ma_fa}
\end{figure*}

In this \emph{paper}, we confront the most recent PTA data for both cases: \emph{i)} $N_{\rm DW} = 1$ where the SGWB in the PTA range dominantly comes from the cosmic strings, and \emph{ii)} $N_{\rm DW} > 1$ where the SGWB in the PTA range comes from the domain walls.
These two cases correspond to axions of two utterly different mass ranges. 
For case \emph{i)}, the cosmic strings live long; that is, $m_a$ is small. Instead, the case \emph{ii)} corresponds to the large $m_a$ region.
We compare the GW spectra in Fig.~\ref{fig:spectrum_gw} for different benchmark points, corresponding to locations in the \{$m_a, F_a$\} plane are shown in Fig.~\ref{fig:ma_fa}.

\begin{figure*}[t]
	\centering
\includegraphics[width=1\linewidth]{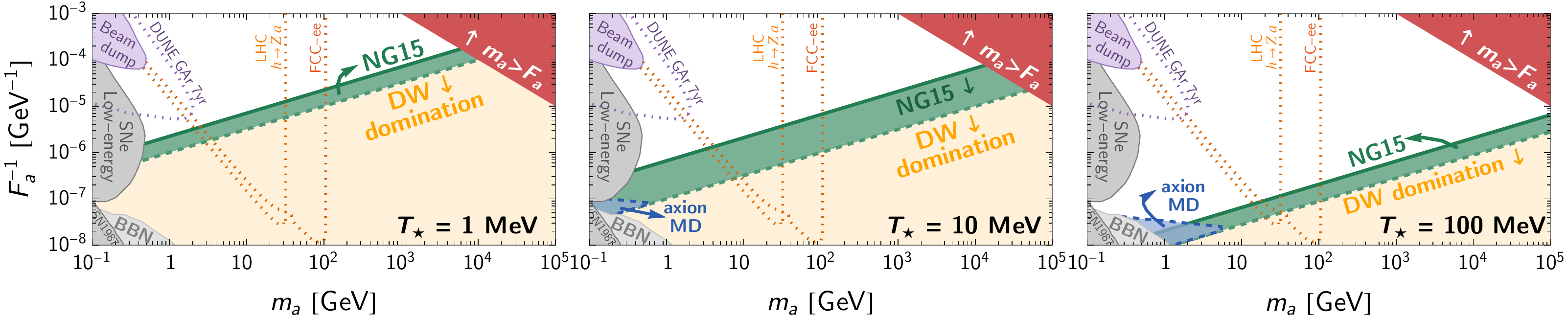}\\[-1em]
	\caption{The PTA-DW constraint (in \textcolor{PineGreen}{\bf green}) changes with $T_\star$. 
For fixed $T_\star$ and $m_a$, the constrained range of $F_a$ in \textcolor{PineGreen}{\bf green} is derived from the $\alpha_\star$ constrained region of Fig.~\ref{fig:main_result}-right, using Eq.~\eqref{eq:DW_alpha}. The \textcolor{YellowOrange}{\bf yellow} region corresponds to $\alpha_\star > 1$, which corresponds to the DW domination and can change the GW prediction; we do not extend the constraint into this region.
 For $T_\star \gtrsim 302$ MeV (\emph{cf.} Fig.~\ref{fig:main_result}-right), NG15 data constrains $\alpha_\star > 1$; that is, the green band overlays part of the yellow region. The \textcolor{RoyalBlue}{\bf blue} region is where the axions -- produced from DW annihilations -- dominate the Universe before they decay prior to  BBN. In this case, the theoretical prediction for the GW spectrum also has to be re-evaluated.
 }
	\label{fig:largemass_constraint}
\end{figure*}

\section{Searching and constraining SGWB with PTA}
\label{sec:data_analysis}
This work analyzes the recent NG15 data set \cite{andreamitridate2023_8102748} covering a period of observation $T_{\rm obs} = 16.03$ years \cite{NANOGrav:2023gor}. From the pulsar timing residuals, the posterior probability distributions of the global-string and domain-wall model parameters are derived.
We consider 14 frequency bins of NG15 data, where the first and last bins are at $1/T_{\rm obs} \simeq 1.98$ nHz  and $14/T_{\rm obs} \simeq 27.7$ nHz, respectively.
The  analysis is done by using \texttt{ENTERPRISE} \cite{enterprise, enterpriseextension} via the handy wrapper \texttt{PTArcade} \cite{andreamitridate2023, Mitridate:2023oar}. The priors for the model parameters are summarized in Tab.~\ref{tab:prior} in Appendix \ref{app:prior}. We refer readers to Ref.~\cite{NANOGrav:2023hvm} for a short review of Bayesian analysis.

This work considers the SGWB in the two  scenarios discussed above together with the astrophysical background. 
Fig.~\ref{fig:main_result}-middle and -right show the 68\%-CL (or $1\sigma$) and 95\%-CL (or $2\sigma$) in dark and light blue regions, respectively.
We obtain the best-fit values $f_a = 9.87^{+2.67}_{-2.02} \times 10^{15}$ GeV and $T_{\rm dec} = 3.50^{+2.44}_{-1.48}$ MeV for global strings, and $\alpha_\star = 0.114^{+0.060}_{-0.033}$ and $T_\star = 128^{+55}_{-33}$ MeV for domain walls.
The global-string and domain-wall SGWB are preferred over the SMBHB signal implemented by \texttt{PTArcade}, as suggested by their Bayes Factors (BF) larger than unity (BF$_{\rm CS} = 26.0$, BF$_{\rm DW} = 44.7$) when compared to the SMBHB interpretation; \emph{cf.} Eq.~(9) of \cite{NANOGrav:2023hvm}. 
 We show the best-fitted spectra for these two new-physics cases in Fig.~\ref{fig:main_result}-left. Translating into axion parameters via Eq.~\eqref{eq:Tdec_ma} and \eqref{eq:DW_alpha}, the best fits correspond to $\{f_a,m_a\} = \{9.87 \times 10^{15} ~ {\rm GeV}, 4.78 \times 10^{-15} ~ {\rm eV}\}$ for global strings (excluded by the axion overabundance) and $m_a F_a^2 = 2.6 \times 10^{15} ~ {\rm GeV}^3$ for domain walls.
For completeness, we show the case without the SMBHB contribution in App.~\ref{app:best_fit}. Because the two new-physics cases explain the data well by themselves, we see that the $1\sigma$ and $2\sigma$ regions of Fig.~\ref{fig:main_result} match those without the SMBHB in Fig.~\ref{fig:pdf_global_string_st}. The values of the best fits, given in App.~\ref{app:best_fit}, only change slightly.
 
Although the two scenarios could explain the signal, this work aims to set bounds on the model parameter space associated with a too strong SGWB in conflict with the NG15 data.
Following \cite{NANOGrav:2023hvm},
we identify excluded regions of the new-physics parameter spaces using the posterior-probability ratio (or $K$-ratio).  
Specifically, the excluded gray regions in Fig.~\ref{fig:main_result}-middle and -right correspond to the areas of parameter spaces where the $K$-ratio between the combined new-physics+SMBHB and the SMBHB-only models drops below 0.1\footnote{\emph{i.e.}, the new-physics contribution makes the overall signal \emph{strongly disfavored} by the data}, according to Jeffrey's scale \cite{jeffreys1998theory}, due to a too-strong SGWB from the new-physics model.
We emphasize that the values of the BFs strongly depend on the modeling of the SMBHB signal as it is the ratio of evidence of the considered model and the SMBHB template. However, the constrained regions depend only slightly on it \cite{NANOGrav:2023hvm}.

We emphasize that the constraints on the axion parameter space presented in this paper are not the same as the regions of best-fit obtained in the literature using the previous dataset, \emph{e.g.,} \cite{Ferreira:2022zzo, Madge:2023cak}.
For fitting the PTA data, a particular part of the GW spectrum is preferred; thus, the best-fits region is allowed within a tight parameter space (the blue blobs in Fig.~\ref{fig:main_result}).
On the other hand,
the constraint can be drawn from any part of the spectrum if the GW signal becomes too large and disfavored by the data. So, the constraint can be extended over a vast parameter space (the grey regions in Fig.~\ref{fig:main_result}).
Now we discuss, in turn, the NG15 constraints -- on global strings ($N_{\rm DW} = 1$) and domain walls ($N_{\rm DW} > 1$) -- and translate them into the constraints in the axion parameter space.

\subsection{Result for $N_{\rm DW} = 1$, implications for light axions}
We fit the PTA data with the global-string SGWB, varying $\{f_a, T_{\rm dec}\}$. The 2D posterior result is shown in Fig.~\ref{fig:main_result}, and the dark-blue region is where the cosmic-string SGWB dominates and fits the data to the significance of 1$\sigma$ with the best fit $\{f_a,m_a\} \simeq \{9.9 \times 10^{15} ~ {\rm GeV}, 4.8 \times 10^{-15} ~ {\rm eV}\}$, shown as the benchmark case {\Large $\star$} in Figs.~\ref{fig:spectrum_gw} and \ref{fig:ma_fa}. Note that this benchmark point is excluded by the axion overabundance constraint [see Eq.~\eqref{eq:dm_axion_string}]. A too-large global-string SGWB is constrained by PTA in the grey region of Fig.~\ref{fig:main_result}-middle.
For small $f_a$, the GW from cosmic strings cannot fit the data as its amplitude becomes too small.

As $T_{\rm dec} \ll 0.1$ MeV ($m_a \ll 10^{-17}$ eV), the cut-off \eqref{eq:freq_cs_axion_mass} associated with $T_{\rm dec}$ moves below the PTA window ($f_{\rm GW}(T_{\rm dec}) <$ nHz).
The constraint in this case, Fig.~\ref{fig:main_result}-middle, reads $f_a < 2.8 \times 10^{15}$ GeV ($m_a$-independent), which is stronger than the LIGO bound\footnote{Derived by solving numerically Eq.~\eqref{eq:amp_cs} with $f_{\rm GW} \simeq 20$ Hz and $h^2\Omega_{\rm GW} \simeq 10^{-8}$ for LIGO.} ($f_a \lesssim 8 \times10^{16}$ GeV).
For completeness, we also analyzed the case of stable global strings (\emph{i.e.,} $m_a \to 0$) in App.~\ref{app:stable_string_fitting}, and we obtained a similar bound.
For $T_{\rm dec} \gg 0.1$ MeV ($m_a \gg 10^{-17}$ eV), the cut-off sits at a frequency higher than the PTA window, and the SGWB signal is dominated by the IR tail signal, which scales as $\Omega_{\rm GW} \propto f_{\rm GW}^3$.
From Eqs.~\eqref{eq:freq_cs} and \eqref{eq:amp_cs}, we obtain the asymptotic behavior of $T_{\rm dec} \propto f_a^{4/3}$ (or $m_a \propto f_a^{8/3}$), up to the log correction in Eq.~\eqref{eq:amp_cs}, toward large $f_a$ limit.
We show this bound (\emph{green-region}) in the usual axion parameter space in the \emph{bottom-left} corner of Fig.~\ref{fig:ma_fa}. 
The NG15 constraint on $f_a$ values for $N_{\rm DW} = 1$ corresponds to $f_a > H_{\rm inf}/(2\pi)$. Therefore, it does not apply to cosmic strings linked to quantum fluctuations during inflation. 

Note that Eqs.~\eqref{eq:freq_cs} and \eqref{eq:amp_cs} assume a standard cosmological history, i.e., a transition between the radiation era and the matter era occurring at  $T_{\rm eq}\sim 1$ eV. In the region of parameter space where the axion abundance from the string network exceeds the dark matter abundance [see Eq.~\eqref{eq:dm_axion_string}], the matter era starts earlier, and the cosmological evolution is not viable.
The non-standard cosmological history will modify the PTA data (e.g., the calibration of pulsar timing data and the dispersion measure) and also the SMBHB modeling \cite{NANOGrav:2023ctt}.
Ignoring its impact on PTA data, we can still estimate how the axion overabundance affects our constraint, just from the dilution effect on the GW spectrum \cite{Chang:2019mza,Gouttenoire:2019kij,Chang:2021afa}; see Eq.~\eqref{eq:modified_cs_spectrum_with_axion_overproduction} in App.~\ref{app:EMD_GW_cosmic_string}. 
In Fig.~\ref{fig:ma_fa}, the dot-dashed green line shows the modified PTA constraint due to the diluted GW spectrum from the axion overabundance; see App.~\ref{app:EMD_GW_cosmic_string} for the estimate of the scaling.

\emph{$\Delta N_{\rm eff}$ \& dark matter constraints.}--Although the PTA constraint excludes a large region of the axion parameter space, there exist other theoretical bounds. Axionic strings are known to emit axion particles dominantly \cite{Davis:1986xc}.
Depending on its mass, the axion can contribute to either dark radiation or cold dark matter.
Axions that are relativistic at the time of Big Bang Nucleosynthesis (BBN) are subject to the dark radiation bound expressed as a bound on the number of extra neutrino species, $\Delta N_{\rm eff} < 0.46$ \cite{Planck:2018vyg}. There are uncertainties in deriving this bound linked to the log-correction to the number of strings in the global-string network evolution \cite{Gorghetto:2018myk,Gorghetto:2020qws,Buschmann:2021sdq, Hindmarsh:2021vih}. In this paper, we quote two bounds: the one relying on the semi-analytic calculation \cite{Chang:2021afa} by Chang and Cui (CC), and the lattice result \cite{Gorghetto:2021fsn} by Gorghetto, Hardy, and Nicolaescu (GHN):
\begin{align}
    f_a \lesssim 10^{15} \, {\rm GeV}\left[\frac{\Delta N_{\rm eff}}{0.46}\right]^{\frac{1}{2}} \times \begin{cases}
    3.5  &{\rm \small (CC)},\\
       0.88 \left[\frac{90}{\log\left(\frac{f_a}{H_{\rm BBN}}\right)}\right]^{3/2} &{\rm \small (GHN)},
    \end{cases}
    \label{eq:neff_axion_string}
\end{align}
where we implicitly assume $\lambda \sim 1$ for the GHN bound and $H_{\rm BBN} \simeq 4.4 \times 10^{-25}$ GeV is the Hubble parameter at BBN scale ($T_{\rm BBN} \simeq$ MeV).
Since ALPs have a small mass at late times, they behave as cold dark matter. Subject to the uncertainty in simulations \cite{Hindmarsh:2019csc, Gorghetto:2021fsn, Buschmann:2021sdq}, the abundance $\Omega_a$ of axion dark matter from strings predicted by GHN sets a constraint on the axion,
\begin{align}
    f_a \lesssim &1.8 \times 10^{15} \, {\rm GeV} \sqrt{\left[\frac{\Omega_{a}}{0.266}\right] \left[\frac{25 \times x_{0,a}}{\xi_* \times 10}\right] \left[\frac{g_*(T_{\rm dec})}{3.5}\right]^{1/4}} \times \nonumber\\
    &~ ~ ~ \times \sqrt{\left[\frac{10^2}{\log(f_a/m_a)}\right] \left[\frac{10^{-22}{\rm eV}}{m_a}\right]^{1/2}},
    \label{eq:dm_axion_string}
\end{align}
typically $\xi_* \approx 25$ and $x_{0,a} \approx 10$ \cite{Gorghetto:2021fsn}.
Note that the collapse of the string-wall system\footnote{The collapse of the system when cosmic strings re-enter the horizon also produces GW \cite{Ge:2023rce} when the string (domain-wall) formation happens before (after) inflation, \emph{e.g.,} in the pre-inflationary axion scenario.} at $H \simeq m_a$  produces an axion abundance of the same order as the one from strings \cite{Gorghetto:2020qws}, therefore an $\mathcal{O}(1)$ correction is expected in $\Omega_a$ in Eq.~\ref{eq:dm_axion_string}.
We show both dark radiation and dark matter bounds in Fig.~\ref{fig:ma_fa}. We see that the PTA constraint becomes competitive with the equivocal $\Delta N_{\rm eff}$ bound for $m_a \lesssim 10^{(-22,-23)}$ eV.

\emph{Effects of non-standard cosmology.}--So far, the standard $\Lambda$CDM cosmology \cite{Planck:2018vyg} has been assumed. On the other hand, alternative expansion histories to the usually assumed radiation era are not unlikely above the BBN scale, such as a period of matter domination or kination resulting in a strongly different spectrum of GW for cosmic strings \cite{Chang:2019mza, Chang:2021afa, Gouttenoire:2019kij, Gouttenoire:2021jhk, Simakachorn:2022yjy}.
Nonetheless, the non-standard cosmology modifies the cosmic-string GW spectrum in the high-frequency direction.
From Eq.~\eqref{eq:freq_cs}, the non-standard era must end below the MeV scale to substantially change the SGWB in the PTA window. We have checked the effects of matter and kination eras with \texttt{PTArcade} and found that   
such SGWB distortion cannot improve the global string interpretation of PTA data. Besides, we expect only a negligible effect on the PTA bound obtained in this work.

\emph{QCD axion.}--From Fig.~\ref{fig:ma_fa}, the PTA data can exclude some parts of the QCD axion (red line). However, this region of parameter space is already excluded due to the overabundance of axion dark matter or due to $\Delta N_{\rm eff}$ bounds. 
To relax these bounds, one can invoke a scenario where cosmic strings decay during a matter-domination era (or any era with the equation-of-state smaller than that of radiation), which efficiently dilutes these relics but still allows for a GW signal in the PTA frequency range \cite{Ramberg:2019dgi, Ramberg:2020oct, Chang:2021afa}. Interestingly, such matter-domination era at early times can imprint a specific signature in the SGWB from global strings, which can be observed in future-planned GW experiments at frequencies above nHz frequencies \cite{Chang:2019mza, Chang:2021afa, Gouttenoire:2019kij, Ghoshal:2023sfa}.

\subsection{Result for $N_{\rm DW} > 1$, implications for heavy axions}
We fit the DW SGWB, varying $\{T_{\star}, \alpha_\star, \beta, \delta\}$, to the PTA data. Because the posteriors of $\beta$ and $\delta$ are unconstrained, we show only the 2D posterior of $\{T_{\star}, \alpha_\star\}$ in Fig.~\ref{fig:main_result}-right. The DW SGWB can fit the PTA data in the dark-blue region to $1\sigma$. The best fit value of $\{T_\star, \alpha_\star\}$ is translated via Eq.~\eqref{eq:DW_alpha} into $m_a F_a^2 \simeq 2.6 \times 10^{15}$ GeV and corresponds to the benchmark spectrum and line in Figs.~\ref{fig:spectrum_gw} and \ref{fig:ma_fa}, respectively. However, for large enough $\alpha_{\star}$, DW generates a GW signal well stronger than the PTA signal, leading to a constraint in the gray region in Fig.\ref{fig:main_result}-right.
The constraint is the strongest $\alpha_{\star} \lesssim 0.02$ at $T_\star \simeq 13.8$ MeV when the peak of the SGWB is centered in the PTA window; see also Eq.~\eqref{eq:freq_dw}.
For $T_\star > 13.8$ MeV ($< 13.8$ MeV), the GW spectrum has its IR (UV) tail in the PTA range; thus, the constraint on $\alpha_\star$ becomes weaker.

For heavy axions with $Z_{N_{\rm DW}}$-symmetry whose mass depends on the explicit-symmetry-breaking scale  $\Lambda \simeq \sqrt{m_a F_a}$ where $F_a = f_a/ N_{\rm DW} $,
the PTA constraint in Fig.~\ref{fig:main_result}-right is translated via Eq.~\eqref{eq:DW_alpha} into a bound on \{$F_a, m_a$\} with the degeneracy among them. For a fixed $T_\star$, we obtain the excluded region on the axion parameter space, \emph{i.e.,} the green region of Fig.~\ref{fig:largemass_constraint}.  
Very large $f_a$ corresponds to $\alpha_\star > 1$; the DW-domination era occurs before it decays and should affect the GW prediction.
We do not extend our PTA bound in the DW domination regime, shown in the yellow of Fig.~\ref{fig:largemass_constraint}. In fact, Eq.~\eqref{eq:GWDWspec} assumes a radiation-dominated Universe. Constraining the DW-domination region requires computing the evolution of the DW network and its SGWB in a Universe with a modified equation of state. We leave this non-trivial task for future investigation; see also \cite{Bai:2023cqj}. To be conservative, we mark this region unconstrained for now, although we expect some constraints will prevail there.

Because the PTA constraint on $\alpha_\star$ is not linear in $T_\star$, the width of the PTA band is maximized only for $T_{\star} \simeq 13.8$ MeV where the bound on $\alpha_\star$ is the strongest.
In Fig.~\ref{fig:ma_fa}, we also show the ability to constrain axion parameter space with the PTA-DW signal.
We obtain the constraint by summing the excluded regions for the range ${\rm MeV} \leq T_{\star} \lesssim 302 \, {\rm MeV}$, where $T_{\star} \simeq 302 \, {\rm MeV}$ is where the constraint has $\alpha_\star \geq 1$ in Fig.~\ref{fig:main_result}-right. The upper limit of the green region (large-$m_a$) of Fig.~\ref{fig:ma_fa} is set by the constraint at $T_\star = {\rm MeV}$: $\alpha_\star \gtrsim 0.2$; see Fig.~\ref{fig:main_result}-right. Using Eq.~\eqref{eq:DW_alpha}, this upper bound is defined as $m_a F_a^2 \gtrsim 2 \times 10^{11} ~ {\rm GeV}^3$. Some regions above and within the green band (smaller $m_a F_a^2$) will be probed by future particle physics experiments  \cite{Bauer:2017ris, Bauer:2018uxu, Hook:2019qoh, Bauer:2021mvw, Alonso-Alvarez:2023wni}.

Other than the PTA bound, the $\{T_{\star}, \alpha_\star\}$ parameter space is subject to theoretical constraints related to the DW decay and its by-products. In this work, we consider that the heavy axion produced from the DW decay subsequently decays into SM particles, \emph{e.g.,} photons via $\mathcal{L} \supset - \frac{g}{4} F \tilde{F} \theta$ with the decay rate $\Gamma_{\theta\gamma} = m_a^3 g^2/(64 \pi)$ \cite{Cadamuro:2011fd}.
Using this to $F_a = 1.92 \alpha_{\rm EM}/ (2 \pi g_{\theta\gamma})$, the decay is efficient when $\Gamma_{\theta\gamma} > H(T)$, which is equivalent to,
\begin{align}
    T < T_{\theta\gamma} \equiv 236 \, {\rm MeV} \left[\frac{10.75}{g_*(T_{\theta\gamma})}\right]^{\frac{1}{4}} \left[\frac{m_a}{\rm GeV}\right]^{\frac{3}{2}} \left[\frac{10^6 \, {\rm GeV}}{f_a/N_{\rm DW}}\right].
\end{align}
The bound $T_{\rm BBN} < T_{\theta \gamma}$ is similar to the BBN bound from \cite{Depta:2020wmr} in Figs.~\ref{fig:ma_fa} and \ref{fig:largemass_constraint}.

Moreover, the heavy axion which behaves non-relativistically might decay after it dominates the Universe if $T_\star > T_{\rm dom} >T_{\theta\gamma}$ where the temperature  $T_{\rm dom}$ corresponds to the heavy-axion domination, \emph{i.e.,} $\rho_{a}(T_{\rm dom}) = \rho_a(T_\star) (a_\star/a_{\rm dom})^3 = \rho_{\rm tot} (T_{\rm dom})$,
\begin{align}
    T_{\rm dom} \simeq 0.02 ~{\rm MeV}\left[\frac{10.75}{g_*(T)}\right]^{\frac{1}{2}} \left[\frac{50 \rm MeV}{T_\star}\right] \left[\frac{m_a}{\rm GeV}\right] \left[\frac{f_a/N_{\rm DW}}{10^6{\rm GeV}}\right]^2.
\end{align}
We mark this region in the blue region of Fig.~\ref{fig:largemass_constraint}.
For the sum of PTA constraints varying $T_\star$ in Fig.~\ref{fig:ma_fa}, we omit showing the color of the axion matter-domination (MD) region, which cuts the PTA region from the low-$m_a$ region\footnote{Using Eq.~\eqref{eq:DW_alpha} with $\alpha_\star = 1$ and $T_{\theta\gamma} < T_{\rm dom}$, the cut follows $236 (m_a/{\rm GeV}) < (F_a/ 10^6 {\rm GeV})^2$.}.
This heavy axion induces a matter-domination era that would change the GW prediction, \emph{e.g.,} the causality tail of the spectrum gets distorted \cite{Hook:2020phx, Racco:2022bwj, Franciolini:2023wjm}.
Although this spectral distortion would change the fitting of the data, it would affect the constraint derived here minimally for two reasons.
First, the blue region in Fig.~\ref{fig:largemass_constraint}, leading to the axion-MD, is smaller than the constrained region.
Second, within this region, we find $T_{\rm dom}/T_{\theta \gamma} \lesssim 10$ which leads to $\Omega_{\rm GW } \propto f_{\rm GW}$ for frequencies in the range  $[10^{-2/3}, 1] \times f_{\rm p}^{\rm DW}$, using Eq.~(4.5) of \cite{Hook:2020phx} where $f_{\rm p}^{\rm DW}$ is the peak frequency \eqref{eq:freq_dw}.

\emph{Other effects.}--The friction from axionic DW  interactions with particles of the thermal plasma could change the network's dynamics \cite{Blasi:2023sej} and potentially the SGWB spectrum.
Another effect that could change the bounds is the potential collapse of DW into primordial black holes \cite{Vachaspati:2017hjw, Ferrer:2018uiu, Sakharov:2021dim, Gelmini:2022nim, Gelmini:2023ngs, Gouttenoire:2023ftk, Guo:2023hyp}. Nonetheless, since the prediction is based on the spherical collapse, we would need a large-scale numerical simulation of DW to check whether the PBH formation can be realized.
Lastly, further QCD effects can impact the DW decays relevant for PTA \cite{Kitajima:2023cek, Bai:2023cqj, Lu:2023mcz}.

\section{Conclusion}
\label{sec:conclude}
We analyzed the consequences of the 15-year NANOGrav data on the parameter space of postinflationary axions. 
The bounds in Fig.~\ref{fig:ma_fa} come in two distinct regimes: the low and large axion mass ranges, which are respectively associated with signals from axionic global strings ($N_{\rm DW} = 1$) and domain walls ($N_{\rm DW} > 1$).
In the low-axion-mass region, the constraint on $f_a $  is strongest for $m_a \ll 10^{-17}$ eV, and reads $f_a < 2.8 \times 10^{15}$ GeV. It is competitive with the $\Delta N_{\rm eff}$ bound.
At high masses, $0.1 \mbox{ GeV}\lesssim m_a\lesssim 10^3$ TeV, a substantial region, corresponding to $m_a (f_a/N_{\rm DW})^2\gtrsim 2 \times 10^{11}$ GeV$^3$, can be excluded for DW decaying in the $T_*\propto \sqrt{V_{\rm bias}}\sim$ $1-300$ MeV range.

This study motivates the investigation of the SGWB in the regime of DW domination, as this knowledge could lead to substantial new constraints at large $m_a$ and $f_a$ values.
Once the network of DW dominates the Universe, the scaling regime might be lost. DW would instead enter the stretching regime \cite{Martins:2016lzc} where the energy density scales as $\rho\propto a^{-1}$, the equation of state of $-2/3$ leading to the accelerated cosmic expansion could be in tension with several cosmological observations \cite{Friedland:2002qs, Bai:2023cqj}. Moreover, a period of early DW domination together with the axion matter domination can also affect the SGWB spectra from DW and cosmic strings \cite{Guedes:2018afo, Gouttenoire:2019kij, Cui:2019kkd, Hook:2020phx, Racco:2022bwj, Franciolini:2023wjm}.

 To conclude, GW is a promising tool to probe axion physics. PTA measurements have opened the possibility of observing the Universe at the MeV scale, enabling us to constrain several classes of axion models. By combining NG15 with other data sets from EPTA, InPTA, PPTA, and CPTA collaborations, the constraints on axions can become more stringent, similar to what has been shown for other cosmological sources \cite{Liu:2023ymk, Figueroa:2023zhu}. 
Other planned GW observatories will permit the search for different parts of the predicted SGWB from axion physics and probe the axion parameter spaces uncharted by the PTA.
 Moreover, the synergy of GW experiments over a wide frequency range will allow us to distinguish the axion-GW signals from other SGWB from astrophysical and cosmological sources \cite{Caprini:2018mtu}.

\section*{Acknowledgement}
We are indebted to Andrea Mitridate for teaching us \texttt{PTArcade} and for his substantial help on the analysis.
We thank Marco Gorghetto for discussions and  Matthias Koschnitzke for his technical support.
PS is funded by Generalitat Valenciana grant PROMETEO/2021/083. This work is supported by the Deutsche Forschungsgemeinschaft under Germany’s Excellence Strategy -- EXC 2121 ,,Quantum Universe`` – 390833306 and the Maxwell computational resources operated at Deutsches Elektronen-Synchrotron (DESY), Hamburg, Germany.


\bibliographystyle{JHEP}
\bibliography{nano_glob_str}


\appendix

\clearpage
\onecolumngrid


\renewcommand{\thepage}{S\arabic{page}}
\setcounter{page}{1}


\begin{center}
\textit{\Large Supplemental Material
}
\end{center}
\noindent

This \emph{supplemental material} gives more details on analyzing NG15 data with the global-string and domain-wall templates. App.~\ref{app:prior} specifies the priors used in this study. 
App.~\ref{app:EMD_GW_cosmic_string} discusses the possible modification of the PTA constraint from global strings in the parameter region where the axion is overabundant  (even though this scenario is excluded).
We then present in App.~\ref{app:best_fit} the best fits without and with the astrophysical background and compare them using the Bayes Factor (BF) method.
App.~\ref{app:stable_string_fitting} presents the results of the global string template in the limit $T_{\rm dec} \to 0$ (or $m_a \to 0$), the so-called \emph{stable} global strings. 
We also summarize, in App.~\ref{app:comparison}, the confidence levels associated with interpretations of the NG15 dataset with the GW signal discussed in this paper, compared to other cosmological backgrounds considered in \cite{NANOGrav:2023hvm}.
Our analysis includes the temperature dependence of the number of relativistic degrees of freedom $g_*$ and $g_{*s}$, taken from Ref.~\cite{Saikawa:2018rcs}.

\section{Priors for analysis}
\label{app:prior}
Tab.~\ref{tab:prior} shows the ranges of priors for the parameters in global-string and domain-wall scenarios used for the Monte Carlo Markov Chain tools. For the SMBHB signal, we use the prior of power-law fitted spectrum, which is translated from the 2D Gaussian distribution in SMBHB parameters, motivated by the simulated SMBHB populations \cite{NANOGrav:2023hfp}  and implemented in \texttt{PTArcade}. The Bayes factors reported for our two new-physics cases depend on the evidence of this SMBHB template.

\begin{table}[h]
\begin{tabular}{clc}
\hline\\[-0.75em]
\textbf{Models} &  \textbf{Parameters}&  \hspace{5em}\textbf{Priors} \hspace{5em} \\[0.25em] \hline \\[-0.75em]
 \hspace{0.5em} {\bf Global strings} \hspace{0.5em} & $f_a ~ [{\rm GeV}]$: $U(1)$ breaking scale & $\log$--uniform:$[10^{15},10^{17}]$ \\[0.25em]
 & $T_{\rm dec} ~ [{\rm GeV}]$ : Temperature when string network decays & $\log$--uniform:$[10^{-8},10]$\\[0.25em]
 & (related to axion mass $m_a$ via Eq.~\eqref{eq:Tdec_ma}) & \\[0.25em] \hline \\[-0.75em]
 {\bf Domain walls} & $\alpha_\star$ : Energy fraction in DWs at decay & $\log$--uniform:$[10^{-2},1]$\\[0.25em]
 & $T_\star ~ [{\rm GeV}]$ :  DW annihilation temperature & $\log$--uniform:$[10^{-3},10]$\\[0.25em]
 & $\delta$ : Width of GW spectrum & uniform:$[1,3]$ \\[0.25em]
 & $\beta$ : Slope of GW spectrum for $f > f_{\rm p}$ & uniform:$[1,3]$ \\[0.25em] \hline
\end{tabular}
\caption{Ranges of priors for global-string  and domain-wall parameters used for the analysis.}
\label{tab:prior}
\end{table}

\section{Axion matter domination in $N_{\rm DW} = 1$ case}
\label{app:EMD_GW_cosmic_string}
The string network with string tension $\mu = \pi f_a^2 \log(\lambda^{1/2}f_a/H)$ during the scaling regime has energy density $\rho_{\rm net} \simeq \mu/t^2 \simeq G\mu \rho_{\rm tot}$, where we omit the $\mathcal{O}(1)$ numerical factors.
At $H(T_{\rm dec}) \sim m_a$, the network decays into axions (each of energy $\sim H \sim m_a$ \cite{Gelmini:2022nim, Davis:1986xc, Yamaguchi:1998gx, Hiramatsu:2010yu}) with energy density $\rho_{\rm net}(T_{\rm dec})$ where $T_{\rm dec}$ in Eq.~\eqref{eq:Tdec_ma}. They red-shift as non-relativistic particles, $\rho_{\rm net}(T) \propto a^{-3}$, and eventually dominates the SM radiation at temperature
\begin{align}
    T_{\rm dom}' \simeq T_{\rm dec} G \mu(T_{\rm dec}) \left[\frac{g_*(T_{\rm dec}) g_{*s}(T_{\rm dom}')}{g_*(T_{\rm dom}') g_{*s}(T_{\rm dec})}\right],
    \label{eq:Tdom_axion_cs_decay}
\end{align}
where we used $a^{-3} \propto g_{*s}(T) T^3$.
The domination before the radiation-matter equality $T_{\rm dom}' > T_{\rm eq} \simeq 0.75 ~{\rm eV}$ leads to dark matter overabundance. This bound is similar to Eq.~\eqref{eq:dm_axion_string} and the gray region denoted ``DM strings" in Fig.~\ref{fig:ma_fa}.
A universe with $T_{\rm dom}' > T_{\rm eq}$ cannot resemble the standard $\Lambda$CDM model.
Below,  we compute the modified GW spectrum from global strings due to the earlier matter era, although we do not use it for analyzing the PTA data which relies on the standard cosmology assumption, for e.g. the calibration of pulsar timing data and the dispersion measure \cite{NANOGrav:2023ctt}.

We set today's time as when the photon temperature matches the CMB observation.
The GW signal emitted with frequency $f_{\rm GW}^{\rm emit}$ at temperature $T(a_{\rm emit})$ has the frequency today [$f_{\rm GW}^{\rm emit}(a_{\rm emit}/a_0)$] which is the same for $\Lambda$CDM and non-$\Lambda$CDM cases, i.e., $\left(a_{\rm emit}/a_{0}\right)= \left(a_{\rm emit}/a_{0}\right)_{\rm \Lambda {\rm CDM}}$.
On the other hand, the emitted GW energy density [$\Omega_{\rm GW} \sim (\rho_{\rm GW}^{\rm emit}/\rho_{\rm tot,0})(a_{\rm emit}/a_0)^4$] gets diluted as $\rho_{\rm tot,0} > \rho_{\rm tot,0}^{\Lambda {\rm CDM}}$ \cite{Gouttenoire:2019kij}.  We define the dilution factor $\Upsilon$ as
\begin{align}
     \Upsilon(m_a, f_a) \equiv \frac{\Omega_{\rm GW,0}}{\Omega_{\rm GW,0}^{\rm \Lambda{\rm CDM}}} = \frac{\rho_{\rm tot,0}^{\Lambda{\rm CDM}}}{\rho_{\rm tot,0}} \simeq 0.2 \left[\frac{g_{*s}(T_{\rm dom}')}{g_{*}(T_{\rm dom}')} \right] \left(\frac{10 ~ \rm eV}{T_{\rm dom}'}\right),
\end{align}
which scales as $\Upsilon \propto m_a^{-1/2} f_a^{-2}$, neglecting the log-correction and using Eqs.~\eqref{eq:Tdec_ma} and \eqref{eq:Tdom_axion_cs_decay}. For $\Upsilon < 1$ the axion is dominating the Universe today.
Due to axion overabundance, the GW spectrum in Eq.~\eqref{eq:amp_cs} becomes
\begin{align}
    \Omega_{\rm GW,0}(f_{\rm GW}) = \Omega_{\rm GW,0}^{\Lambda {\rm CDM}}(f_{\rm GW}) \Upsilon(m_a,f_a) \times \mathcal{F}(f_{\rm GW}, f_{\rm GW}^{\rm dom}),
    \label{eq:modified_cs_spectrum_with_axion_overproduction}
\end{align}
where the shape function $\mathcal{F}$ represents the modified causality tail due to the matter domination \cite{Hook:2019qoh} below the horizon-scale frequency at the start of matter domination, i.e., $\Omega_{\rm GW} \propto f_{\rm GW}$ for $f_{\rm GW} < f_{\rm GW}^{\rm dom} = H_{\rm dom} (a_{\rm dom}/a_0)$ instead of $\Omega_{\rm GW} \propto f_{\rm GW}^{3}$ during radiation era. 

Assuming the PTA data does not change with the modified cosmology, we use Eqs.~\eqref{eq:amp_cs}, \eqref{eq:Tdec_ma}, \eqref{eq:Tdom_axion_cs_decay} and \eqref{eq:modified_cs_spectrum_with_axion_overproduction} to estimate how the PTA constraint from global strings (the green region in the bottom-left corner of Fig.~\ref{fig:ma_fa}) is deformed due to the axion overabundance.
For $m_a \lesssim 10^{-22} ~ {\rm eV}$, the PTA constraint is compatible with standard cosmology.
For $10^{-22} ~ {\rm eV} \lesssim m_a \lesssim 10^{-17} ~ {\rm eV}$, the GW amplitude gets diluted by the axion overabundance. The constraint scales as $m_a \propto f_a^4$, as opposed to $f_a=$ constant when assuming a standard cosmological evolution. For $m_a \gg 10^{-17}~ {\rm eV}$, the IR tail is constrained by PTA. The  constraint scales asymptotically as $m_a \propto f_a^2$, using the IR tail $\Omega_{\rm GW} \propto f_{\rm GW}$.
We show the modified constraint as the dashed green curve in Fig.~\ref{fig:ma_fa}.

\section{Global-String and Domain-Wall signals  without SMBHB background}
\label{app:best_fit}

\begin{figure}[h!] 
	\centering
	\underline{No astrophysical background from SMBHBs}\\
 	\includegraphics[width=0.45\linewidth]{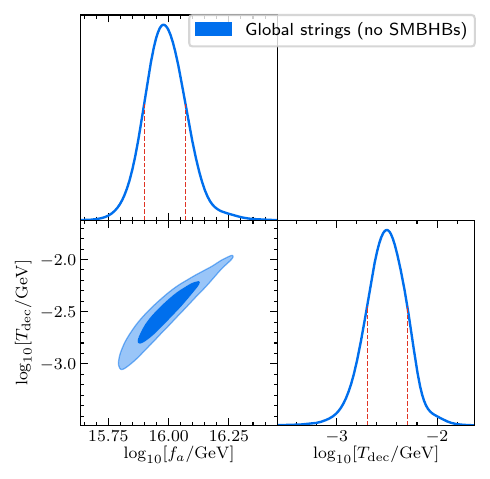}\hfill\vline\hfill
    \includegraphics[width=0.45\linewidth]{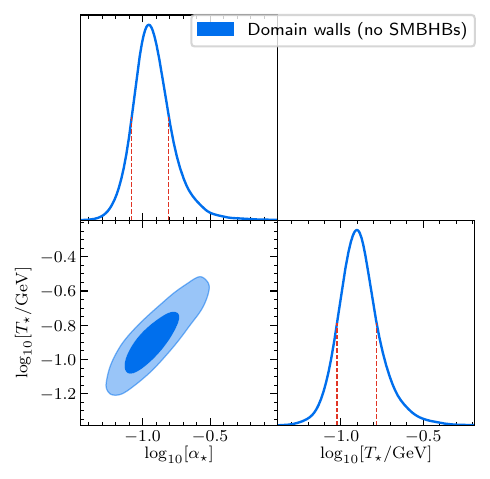}
	\caption{Best fits to NG15 data.  \emph{Left:} The 2D posterior for the global-string SGWB template presented in the main text. Via Eq.~\eqref{eq:Tdec_ma}, the best-fit corresponds to axion parameters $\{f_a,m_a\} = \{9.55 \times 10^{15} ~ {\rm GeV}, 3.89 \times 10^{-15} ~ {\rm eV}\}$. The comparison of the fit to the SMBHB signal yields the BF $\simeq 22.8$. 
	\emph{Right:} Result for domain-wall SGWB, which has the BF $\simeq 23.4$. The best-fitted axion parameters satisfy $m_a F_a^2 = 2.4 \times 10^{15} ~ {\rm GeV}^3$; \emph{cf.} Eq.~\eqref{eq:DW_alpha}. The  posteriors for the UV slope $\beta$ and the width $\delta$ are not constrained as only the IR tail of the spectrum (\ref{eq:spectralshape}) lies within the PTA frequency range for the chosen range of $T_*$.}
\label{fig:pdf_global_string_st}
\end{figure}

In contrast with the analysis presented in the main text, which interprets the NG15 signal in terms of SMBHBs, this appendix assumes the absence of an astrophysical background and instead interpret the signal as a SGWB from global strings or domain walls. Fig.~\ref{fig:pdf_global_string_st} shows the 2-dimensional posterior of the global-string and the domain-wall parameters. For global strings, the best-fit (max. posterior) is at $f_a = 9.55^{+2.19}_{-1.63} \times 10^{15}$ GeV and $T_{\rm dec} = 3.16^{+1.88}_{-1.15}$ MeV at 68\% CL. The central value of $T_{\rm dec}$ correspond to 
$m_a = 3.89 \times 10^{-15} ~ {\rm eV}$; \emph{cf.} Eq.~\eqref{eq:Tdec_ma}.
For domain walls, the best-fit is at $\alpha_\star = 0.111^{+0.045}_{-0.027}$ and $T_\star = 125^{+31}_{-39}$ MeV, with the error within the 68\% CL region.
Their central values give $m_a F_a^2 = 2.4 \times 10^{15} ~ {\rm GeV}^3$; \emph{cf.} Eq.~\eqref{eq:DW_alpha}.
We calculate the Bayes Factor (compared to the SGWB from SMBHBs) from \texttt{PTArcade} and find that the BFs are $22.8$ for global strings and $23.4$ for domain walls.
When the SMBHB background is added, we find that the BF for both cases increases to 26.0 for global strings and 44.7 for domain walls.
However, the values of the best-fitted parameters change only slightly:  $f_a = 9.87^{+2.67}_{-2.02} \times 10^{15}$ GeV and $T_{\rm dec} = 3.50^{+2.44}_{-1.48}$ MeV, at 68\% CL for global strings, corresponding to $m_a = 4.78 \times 10^{-15} ~ {\rm eV}$.
For domain walls, we have $\alpha_\star = 0.114^{+0.060}_{-0.033}$ and $T_\star = 128^{+55}_{-33}$ MeV, corresponding to $m_a F_a^2 = 2.6 \times 10^{15} ~ {\rm GeV}^3$.


\section{Global strings for $m_a \to 0$}
\label{app:stable_string_fitting}
\begin{figure}[h] 
	\centering
	\underline{Global strings in the limit $m_a \to 0$}\\
 	\includegraphics[width=0.325\textwidth]{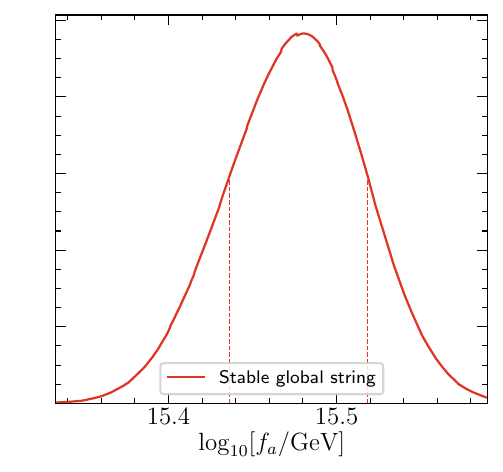}\hfill
    \includegraphics[width=0.33\linewidth]{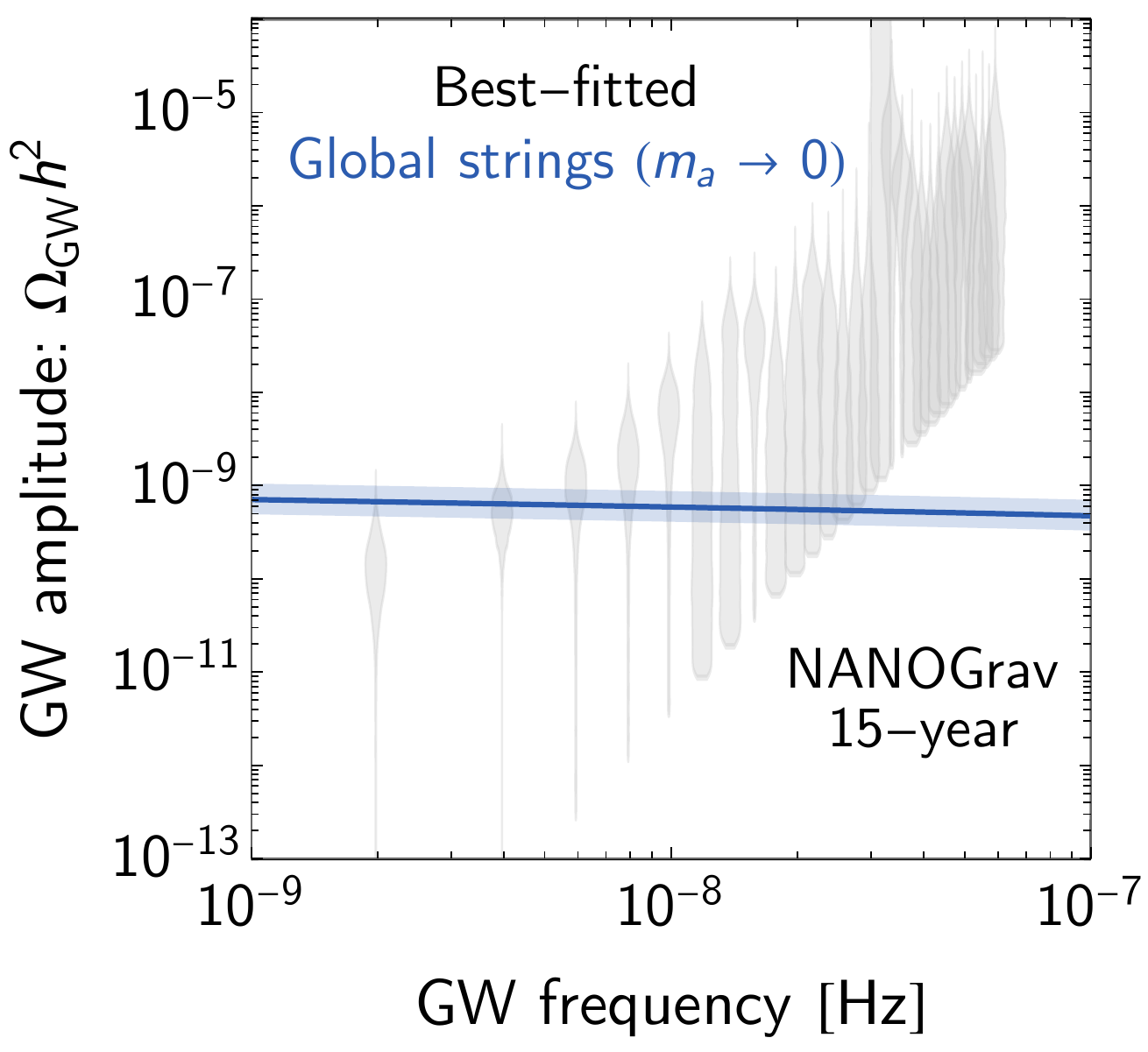}
    \includegraphics[width=0.325\linewidth]{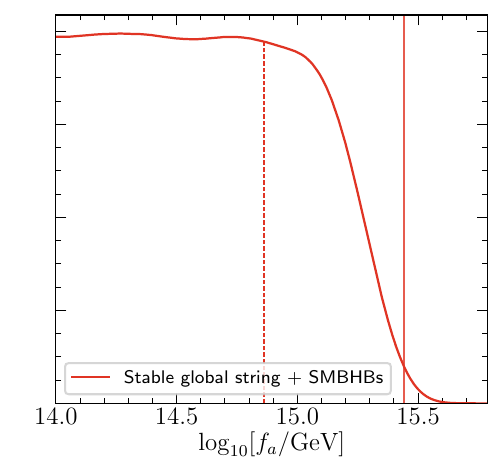}
	\caption{\emph{Left:} The 1D posterior of the stable global-string SGWB, using NG15 data set. The best-fitted  $f_a$ value is $f_a \simeq 2.99^{+0.31}_{-0.26} \times 10^{15} \, \rm GeV$ at 68\% CL and the BF is $1.45 \times 10^{-3}$, for the comparison with the SMBHBs. The vertical red line indicates the 1-$\sigma$ region. \emph{Middle:} The best-fitted GW background from stable global strings and its range within  $1\sigma$ region of $f_a$, laying over the violins of NG15 observation. \emph{Right:} The 1D posterior of the stable global-string SGWB + SMBHBs contribution, fitted to NG15 data set. The best-fitted string scale is $f_a \simeq 1.83^{+5.45} \times 10^{15} \, \rm GeV$ at 68\% CL and the BF of $0.64$, compared to the SMBHBs alone. The vertical dashed line locates the $1\sigma$ region, while the solid vertical line marks the $K$-ratio $=0.1$ and sets a limit on $f_a < 2.77 \times 10^{15}$ GeV.}
 \label{fig:pdf_global_string_st2}
\end{figure}

The constrained region in Fig.~\ref{fig:main_result}-middle shows that the PTA signal from global strings with small $T_{\rm dec}$ (or small $m_a$) reaches the asymptotic value of $f_a \simeq 2.8 \times 10^{15}$ GeV.
This is because the cut-off specified by $T_{\rm dec}$ moves outside of the PTA range, and the SGWB spectrum is seen as the one from stable global strings in the limit $T_{\rm dec}$ or $ m_a \to 0$. 
Fig.~\ref{fig:pdf_global_string_st2}-left shows the 1D posterior of signal from the stable global strings, which has the best-fitted spectrum at $f_a \simeq 2.99^{+0.31}_{-0.26} \times 10^{15} \, \rm GeV$ at 68\% CL. Nonetheless, it has the BF of $1.45 \times 10^{-3}$ due to its red-tilted spectrum, poorly fitting the data, as shown in  Fig.~\ref{fig:pdf_global_string_st2}-middle.
When the SMBHB background is added in Fig.~\ref{fig:pdf_global_string_st2}-right, the BF becomes 0.64, meaning that the stable string spectrum worsens the fit compared to the SMBHB alone. Although the fit is not good, the constraint can be derived when the global-string SGWB becomes too strong (too large $f_a$) using the $K$-ratio, discussed in the main text (see also Ref.~\cite{NANOGrav:2023hvm}). 
The vertical solid line in  Fig.~\ref{fig:pdf_global_string_st2}-right shows the limit set by the NG15 data ($K$-ratio $= 0.1$): $f_a < 2.77 \times 10^{15}$ GeV, which is similar to bound obtain from Fig.~\ref{fig:main_result}-middle in the $T_{\rm dec} \to 0$ limit.

\section{Comparison to other new-physics interpretation of the signal}
\label{app:comparison}
\begin{figure}[h] 
	\centering
\includegraphics[width=\linewidth]{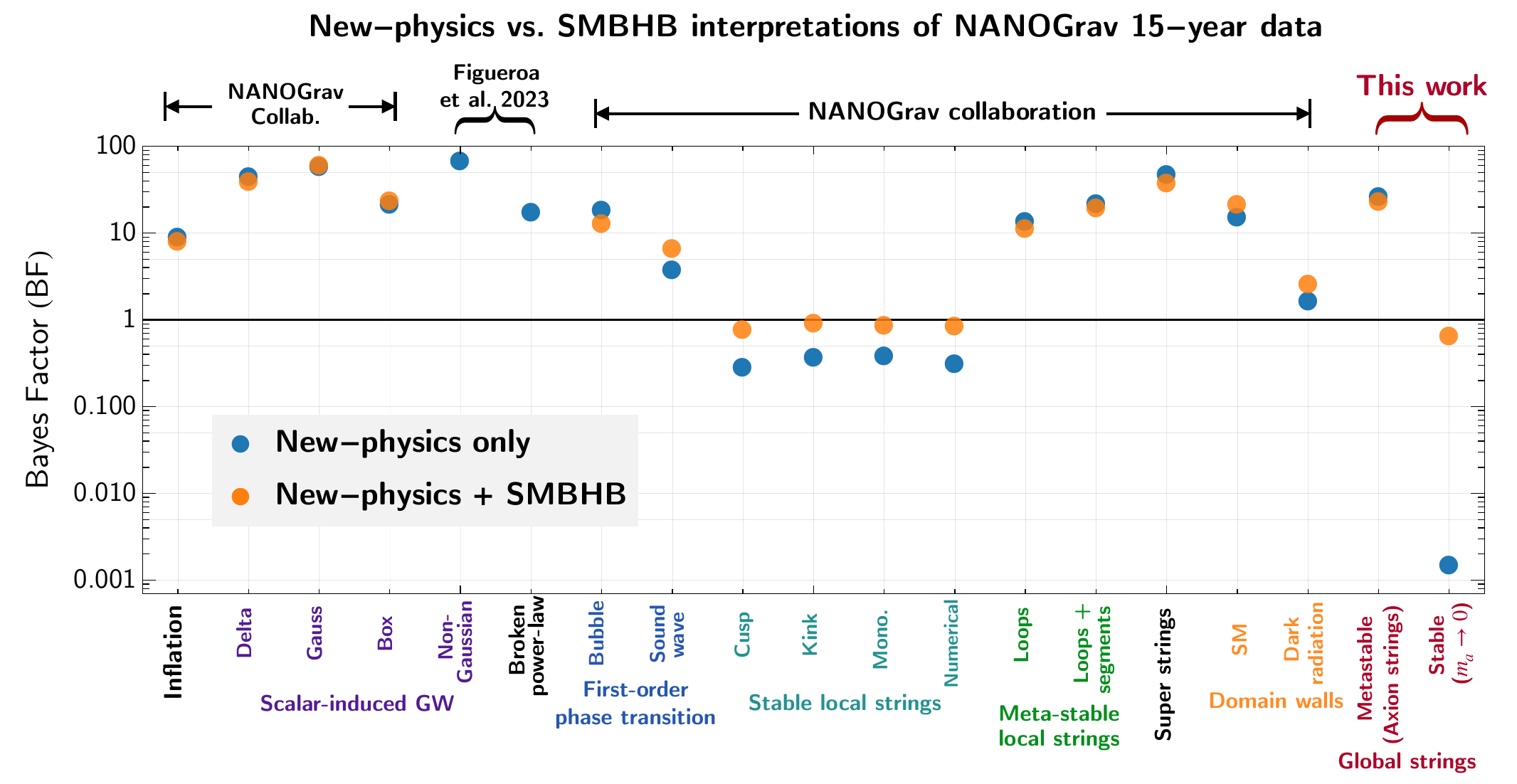}
	\caption{Comparison of the model considered in this work and other new-physics interpretations considered by NANOGrav Collaboration \cite{NANOGrav:2023hvm} and Figueroa et al. \cite{Figueroa:2023zhu}. We only consider the results using the same assumption on SMBHB background \cite{NANOGrav:2023hfp}. This figure extends Fig.~2 of Ref.~\cite{NANOGrav:2023hvm}.}
 \label{fig:model_comparison}
\end{figure}

Fig.~\ref{fig:model_comparison} summarizes confidence levels -- in terms of the Bayes Factor (BF) -- for explaining the NG15 dataset with new-physics interpretations.
We only consider the result from the analysis using the same assumption on the SMBHB background \cite{NANOGrav:2023hfp}.
We also omit our DW result here, which is the same analysis as in  \cite{NANOGrav:2023hvm} and yields similar BFs. Although the axion-string template fits the NG15 well, the best-fit parameter space conflicts strongly with the $\Delta N_{\rm eff}$ and DM abundance constraints, i.e., the benchmark point of the best fit $\star$ sits deep inside the constrained region in Fig.~\ref{fig:ma_fa}.

\end{document}